\documentclass{aa}

\usepackage{mathtools}
\usepackage{nicefrac}

\usepackage{siunitx}
\usepackage{booktabs}
\usepackage{rotating}
\usepackage[version=4]{mhchem}
\usepackage{makecell}
\usepackage{placeins}
\usepackage{csquotes}

\usepackage{hyperref}
\hypersetup{unicode=true, pdftoolbar=true, pdfmenubar=true, pdffitwindow=false, pdfstartview={FitH}, pdftitle={ExoMDN: Rapid characterization of exoplanet interior structures with Mixture Density Networks}, pdfauthor={Philipp Baumeister, Nicola Tosi},	pdfsubject={},	pdfcreator={\LaTeX\ with package \flqq hyperref\frqq}, pdfproducer={pdfTeX \the\pdftexversion.\pdftexrevision}, pdfkeywords={exoplanets, machine learning, neural networks, exomdn, planet interior}, pdfnewwindow=true, colorlinks=true,linkcolor=black,citecolor=black,filecolor=magenta,urlcolor=black, pdfborder= 0 0 0}

\renewcommand{\vec}[1]{\mathbf{#1}}
\newcommand{\dr}[1]{\frac{\mathrm{d} #1}{\mathrm{d} r}}

\newcommand*\subtxt[1]{_{\textnormal{#1}}}
\DeclareRobustCommand\_{\ifmmode\expandafter\subtxt\else\textunderscore\fi}

\DeclareSIUnit{\Me}{M_\oplus}
\DeclareSIUnit{\Re}{R_\oplus}

\graphicspath{{./}{figures/}}

\begin{document}

\title{ExoMDN: Rapid characterization of exoplanet interior structures with Mixture Density Networks\thanks{ExoMDN is freely accessible through the GitHub repository \url{https://github.com/philippbaumeister/ExoMDN}}}

\titlerunning{ExoMDN: Rapid characterization of exoplanet interiors}
\authorrunning{Baumeister \& Tosi}

   \author{Philipp Baumeister \inst{1,2}
          \and
          Nicola Tosi \inst{1}}

   \institute{Institute of Planetary Research, German Aerospace Center (DLR), Rutherfordstraße 2, D-12489 Berlin, Germany\\
              \email{philipp.baumeister@dlr.de}
              \and
              Department of Astronomy and Astrophysics, Technische Universität Berlin, Hardenbergstraße 36, D-10623 Berlin,  Germany}

\date{Received February 22, 2023 / Accepted June 14, 2023}

  \abstract
   {}
   {Characterizing the interior structure of exoplanets is essential for understanding their diversity, formation, and evolution. As the interior of exoplanets is inaccessible to observations, an inverse problem must be solved, where numerical structure models need to conform to observable parameters such as mass and radius. This is a highly degenerate problem whose solution often relies on computationally-expensive and time-consuming inference methods such as Markov Chain Monte Carlo.}
   {We present ExoMDN, a machine-learning model for the interior characterization of exoplanets based on Mixture Density Networks (MDN). The model is trained on a large dataset of more than 5.6 million synthetic planets below 25 Earth masses consisting of an iron core, a silicate mantle, a water and high-pressure ice layer, and a H/He atmosphere. We employ log-ratio transformations to convert the interior structure data into a form that the MDN can easily handle.}
   {Given mass, radius, and equilibrium temperature, we show that ExoMDN can deliver a full posterior distribution of mass fractions and thicknesses of each planetary layer in under a second on a standard Intel i5 CPU. Observational uncertainties can be easily accounted for through repeated predictions from within the uncertainties. We use ExoMDN to characterize the interior of 22 confirmed exoplanets with mass and radius uncertainties below 10\% and 5\% respectively, including the well studied GJ 1214 b, GJ 486 b, and the TRAPPIST-1 planets. We discuss the inclusion of the fluid Love number $k_2$ as an additional (potential) observable showing how it can significantly reduce the degeneracy of interior structures.  Utilizing the fast predictions of ExoMDN, we show that measuring $k_2$ with an accuracy of 10\% can constrain the thickness of core and mantle of an Earth analog to $\approx13\%$ of the true values.}
   {}


   \maketitle
%


\section{Introduction} \label{sec:intro}

In the past decade, the number of discovered exoplanets has been growing rapidly, with more than 5000 planets confirmed to date. Characterizing the interior structures of these planets, that is the size and mass of their main compositional reservoirs, is critical to understanding the processes that govern their formation, evolution, and potential to support life \citep{Spiegel12622, vanhoolst2019ExoplanetInteriors}. 
Numerical models are commonly used to compute interior structures that fit to observed mass and radius of the planet \citep[e.g.][]{sotin2007MassradiusCurve, valencia2007DetailedModels, fortney2007PlanetaryRadii, wagner2011InteriorStructure, zeng2013DetailedModel, unterborn2019PressureTemperature, baumeister2020MachinelearningInference, huang2022MAGRATHEAOpensource}. However, unlike planets in the Solar System for which a wealth of observational data is available ranging from geodetic observations to in-situ seismic measurements, for exoplanets, mass and radius are often the only parameters that can be determined. As a result, the interior structure is highly degenerate, with many qualitatively different interior compositions that can match the observations equally well \citep{rogers2010FrameworkQuantifying, dorn2015CanWe, dorn2017GeneralizedBayesian}. 
Probabilistic inference methods, such as Markov Chain Monte Carlo (MCMC) sampling, are regularly utilized to obtain a comprehensive picture of possible planetary interiors, while also taking into account observational uncertainties \citep{rogers2010FrameworkQuantifying, dorn2015CanWe, dorn2017GeneralizedBayesian, brugger2017ConstraintsSuperEarth, dorn2018SecondaryAtmospheres}. Given prior estimations of interior parameters, probabilistic inference methods allow determining the posterior probabilities that best fit the observations. However, in general MCMC methods are computationally intensive and time-consuming, requiring the calculation of hundreds of thousands of interior structure models. The interior inference of a single exoplanet can therefore take from hours to days. Furthermore, a dedicated framework combining both a forward interior structure model and an MCMC scheme is necessary, which can limit the large-scale applicability of these techniques due to the need for specialized expertise in planetary interior modeling. To fully exploit the ever-increasing number of exoplanet detections, a fast alternative to MCMC inference is therefore needed.

We present here ExoMDN, a standalone, machine-learning-based model which is capable of providing a full inference of the interior structure of low-mass exoplanets in under a second without the need for a dedicated interior model. We make available in a GitHub repository\footnote{\url{https://github.com/philippbaumeister/ExoMDN}} both the trained models and the training routines. The purpose of ExoMDN is to provide a rapid first general characterization of an exoplanet interior, which can then be investigated further with more detailed, specialized models.

\section{Machine Learning for interior characterization} \label{sec:MLintro}

In recent years, machine-learning based methods have become increasingly relevant in planetary science because of their ability to facilitate and speed up otherwise very time-consuming calculations. Deep Neural Networks in particular have been applied to the detection of transits \citep{chaushev2019ClassifyingExoplanet, malik2021ExoplanetDetection, valizadegan2022ExoMinerHighly}, for atmospheric retrievals \citep{marquez-neila2018SupervisedMachine, zingales2018ExoGANRetrieving, himes2022AccurateMachinelearning}, in geodynamic simulations \citep{atkins2016UsingPattern, agarwal2021DeepLearning, agarwal2021ConstrainingMars}, in planet formation models \citep{alibert2019UsingDeep, cambioni2019RealisticOnthefly, emsenhuber2020RealisticOnthefly, auddy2022UsingBayesian}, as well as for characterizing exoplanet interiors \citep{baumeister2020MachinelearningInference, zhao2021MachineLearning, haldemann2023ExoplanetCharacterization}.

In an earlier work \citep{baumeister2020MachinelearningInference}, we presented a proof-of-concept method to characterize exoplanet interiors using Mixture Density Networks (MDNs) \citep{bishop1994MixtureDensity}. MDNs can predict the full probability distribution of parameters by approximating these with a linear combination of Gaussian kernels. We trained an MDN to infer the range of plausible thicknesses of compositional layers in a planet based on mass and radius inputs. However, this was not a full characterization of the interior, as our network could only predict the marginals of the posterior distribution. While this gives an accurate estimation of the range of admissible parameter values, it does not allow pinpointing specific interior structures that fit observed mass and radius nor determining correlations between the various layers. For this, the prediction of the full, multidimensional posterior distribution is required.

ExoMDN builds upon our previous work and is capable of providing a full inference of the entire posterior distribution of interior structures for a planet in a fraction of a second (e.g., on a standard Intel i5 CPU). In addition, we include the equilibrium temperature of the planet as an input parameter to the network in addition to mass and radius, by improving on the atmosphere and water layers in the underlying forward model used to generate the training data. In particular, we use the full water phase diagram compiled by \citet{haldemann2020AQUACollection} in place of the previous simple isothermal, high-pressure ice layer, and model an isothermal atmosphere instead of the previous zero-temperature approach. The use of the equilibrium temperature thus implicitly includes the orbital distance as an observable parameter. We further improve the robustness of the underlying forward model at high pressures by adopting updated high-pressure equations of state (EoS) for the silicate mantle and iron core. A comparison of the old and new forward models can be found in Fig. \ref{fig:modelcomp} in the Appendix.

We use our interior model to first generate a dataset of $\approx$5.6 million synthetic planets spanning the desired parameter space of interior structures, planet masses, and equilibrium temperatures. We then train a Mixture Density Network to predict the parameters of a mixture of multivariate normal distributions, with the aim of approximating the posterior distribution for a given set of input parameters, namely  mass, radius, and equilibrium temperature. In order for the MDN to handle multidimensional predictions, we apply log-ratio transformations on the training data to convert the interior structures into new coordinates that the MDN can easily handle. We present two trained models: One trained on planet mass, radius, and equilibrium temperature; and a second including the fluid Love number $k_2$ as an additional input. Fluid Love numbers describe the shape of a rotating planet in hydrostatic equilibrium. The second-degree Love number $k_2$ is particularly interesting for exoplanet interior characterization, as it depends solely on the interior density distribution \citep{kellermann2018InteriorStructure, padovan2018MatrixpropagatorApproach,baumeister2020MachinelearningInference}. In a body with $k_2=0$, the entire mass is concentrated in the center, while $k_2=1.5$ corresponds to a fully homogeneous body. For a number of exoplanets, $k_2$ is potentially measurable through either second-order effects on the shape of the transit light curve \citep{hellard2019RetrievalFluid, akinsanmi2019DetectabilityShape}, or through the apsidal precession of the orbit \citep{csizmadia2019EstimateLove}.

\section{Methods} \label{sec:methods}

\subsection{Interior model}

We compute planetary interior structures with our code TATOOINE \citep{baumeister2020MachinelearningInference, mackenzie2022EffectImproved}. Each planet consists of compositionally distinct layers. The model takes as input the planet mass $M\_p$, the mass fractions of each layer $w_i$, and the equilibrium temperature $T\_{eq}$ (defined at the top of the atmosphere). From the top of the planet toward the center, the model calculates radial profiles of mass $m$, pressure $P$, and density $\rho$ by solving the equations for mass conservation \eqref{eq:mass_conservation}, hydrostatic equilibrium \eqref{eq:hydrostatic}, as well as the equation of state (EoS, \ref{eq:eos}) relating pressure, density, temperature $T$ and composition $c$:
\begin{subequations}
    \begin{equation}
    \dr{m(r)} = 4 \pi r^2 \rho(r),
    \label{eq:mass_conservation}
    \end{equation}
    
    \begin{equation}
    \dr{P(r)} = - \frac{G m(r) \rho(r)}{r^2},
    \label{eq:hydrostatic}
    \end{equation}
    
    \begin{equation}
    P(r) = f\left(\rho(r), T(r), c(r)\right),
    \label{eq:eos}
    \end{equation}
\end{subequations}
where $G$ is the gravitational constant. The planet radius $R\_p$ is iteratively adjusted until the mass at the planet center approaches zero. This yields a final planet radius and the radius fractions of each layer $d_i$.
We fix the pressure at the top of the atmosphere to \SI{10}{\milli bar}. We focus here on planets below \SI{25}{\Me}.
We consider four distinct layers: An iron core, a silicate mantle, a water/ice layer, and a H/He atmosphere.

\subsubsection{Iron core}
We assume that the core consists of pure, solid, hcp-iron. We use the temperature-dependent, high-pressure EoS by \citet{bouchet2013InitioEquation} for pressures below \SI{234.4}{\giga \Pa}. At higher pressures, we switch to the high-pressure EoS from \citet{hakim2018NewInitio}, valid up to \SI{10}{\tera \Pa}.

The presence of lighter elements in the core such as sulfur or hydrogen can significantly reduce the density of the core, which in turn can have large effects on the core size and consequently on the planet radius \citep{hakim2018NewInitio}. The amount of lighter elements in an exoplanet's core is hard to constrain, as it not only depends on the initial abundances in the protoplanetary disk, but also on the processes of core formation and magma ocean cooling  \citep{hirose2021LightElements}.
For a proper treatment of the interior inference, the amount of light elements should be taken as a free parameter, which will increase the degeneracy of interior structures even more. For simplicity and to better illustrate our method, here we neglect the presence of lighter elements in the core, following an approach commonly used in the exoplanet community \citep[e.g.][]{seager2007MassRadiusRelationships, wagner2011InteriorStructure}. Nevertheless, we note that the uncertainty in the light elements budget of the core can be easily incorporated into our method by sampling from a range of core compositions upon creating the training dataset (Sec. \ref{sec:training_data}).

\subsubsection{Silicate mantle}
The silicate layer consists of an upper mantle composed of olivine \ce{(Mg,Fe)2SiO4} and pyroxene \ce{(Mg,Fe)2Si2O6}, a lower mantle composed of magnesiowüstite \ce{(Mg,Fe)O} and bridgmanite \ce{(Mg,Fe)SiO3}, and a high-pressure magnesiowüstite/post-perovskite phase. The transition from upper to lower mantle is assumed to occur at a fixed pressure of \SI{23}{GPa}, and the transition to post-perovskite at a pressure
\begin{equation}
    P(T) = \SI{89.184}{GPa} + \SI{13.3}{\mega\Pa\per\K} \: T
\end{equation}
after \citet{tateno2009DeterminationPostperovskite}, where $T$ is the adiabatic temperature in the mantle. We model the upper and lower mantle with a modified Tait EoS from \citet{holland2011ImprovedExtended}, and the post-perovskite with the generalized Rydberg EoS described in \citet{wagner2011InteriorStructure}.

Similar to the composition of the core, the mantle composition could be varied, for example using stellar abundances as proxies of planet composition \citep[e.g.,][]{dorn2017BayesianAnalysis, hinkel2018StarPlanet}. Since however the relation between star and planet composition is not straightforward \citep{plotnykov2020ChemicalFingerprints}, for simplicity we assume the silicate mantle to have an Earth-like composition with a molar Mg/Si ratio of 1.131 and a magnesium number (Mg\#) of 0.9 \citep{sotin2007MassradiusCurve}. The Mg/Si ratio determines the mixing ratio of the respective mantle minerals, with the Mg\# determining the ratio of the respective Mg and Fe end members. 

\subsubsection{Water layer}
For the water layer we use the tabulated AQUA EoS from \citet{haldemann2020AQUACollection}, spanning a wide temperature and pressure range and including gas, liquid and solid water phases. Liquid and solid layers are assumed to be fully convective, with an adiabatic temperature profile calculated from the adiabatic gradient given by the AQUA table for any given temperature and pressure. Water vapor is assumed to be part of an isothermal atmosphere at the equilibrium temperature.

\subsubsection{H/He atmosphere}
The low densities of many exoplanets hint at extended primordial envelopes composed of hydrogen and helium \citep[e.g.][]{jontof-hutter2019CompositionalDiversity}. We therefore include an outer gaseous H/He envelope of solar-like composition (71\% hydrogen, 29\% helium by mass) based on the EoS from \citet{saumon1995EquationState}. We treat the atmosphere as isothermal with a temperature equal to the equilibrium temperature, an approach also employed by for example \citet{dorn2017GeneralizedBayesian} and  \citet{zeng2019GrowthModel}. While an isothermal atmosphere certainly does not capture the full complexities of exoplanet atmospheres, more detailed atmosphere models would require the inclusion of additional parameters such as infrared and optical opacities, and the intrinsic temperature of the planet \citep[e.g.][]{guillot2010RadiativeEquilibrium}. Since the goal of this work is to explore the machine-learning method and its applications, we have chosen a simplified atmosphere model to limit the overall model complexity of the model and of the training data. However, more complex atmosphere models, in particular those specifically designed to treat gas giant planets \citep[e.g.,][]{fortney2007PlanetaryRadii, nettelmann2011ThermalEvolution, leconte2012NewVision}, can be easily incorporated into ExoMDN by producing suitable training data.

\subsection{Compositional data}
Ideally, an inference model for planets should provide a set of desired parameters which fully describe the interior, such as the thickness or mass of each interior layer. These quantities represent a type of \textit{compositional data}, where the sum of the $D$ components is always constant (e.g., the planet radius or mass). In our case, we are interested in the relative mass and thickness of each layer in the planet, so that
\begin{equation} \label{eq:compositional}
    \sum_i^{D} x_i = 1,
\end{equation}
where $x_i$ is the relative thickness or mass of the $i$th planet layer. This restricted space is known as the simplex $\mathcal{S}^D$, and is commonly represented in the form of a ternary diagram \citep[e.g.][]{rogers2010FrameworkQuantifying}.

The nature of compositional data can make the statistical treatment cumbersome. The constraints imposed by Eq. \eqref{eq:compositional} can give rise to spurious correlations. The shapes of probability distributions can be distorted and skewed, and trying to fit distributions to sample data may lead to points lying outside the simplex \citep{aitchison1982StatisticalAnalysis, pawlowsky-glahn2006CompositionalData}. In particular, this means that Gaussian distributions, which are commonly used to represent continuous data, can not be utilized directly to describe distributions of compositional data, as parts of the distribution would fall outside the closed space. This point is especially relevant for this work: The simple parameterization of Gaussian distributions makes them convenient candidates for components in mixture distributions in order to approximate arbitrary posterior distribution with neural networks, as the entire mixture is described by only a few parameters. It is therefore highly desirable to extend the usefulness of Gaussian mixtures to the analysis of compositional data, while retaining the convenience of their simple parameterization.

A solution is to introduce a set of coordinate changes called log-ratio transformations \citep{aitchison1982StatisticalAnalysis}, which transforms the data coordinates from the simplex into (unconstrained) real space by way of logarithmic ratios between coordinates. We focus here on the additive log-ratio transformation $\mathrm{alr}: \mathcal{S}^D \rightarrow \mathbb{R}^{D-1}$, which takes the logarithm of pairwise ratios between $D-1$ coordinates and an arbitrarily chosen $D$th coordinate ($x_D$), thereby reducing the dimension of the new space by one: 
\begin{equation} \label{eq:alr}
    \mathrm{alr}(x_i) = y_i = \ln \frac{x_i}{x_D}.
\end{equation}

The back-transformation onto the simplex is given by \citep{aitchison1982StatisticalAnalysis}
\begin{equation} \label{eq:inv_alr}
    \mathrm{alr}^{-1}(y_i) = x_i = 
    \begin{dcases}
        \frac{\exp(y_i)}{1 + \sum_j^{D} \exp(y_j)} & (i = 1,\dotsc,D - 1)\\
        \frac{1}{1 + \sum_j^{D} \exp(y_j)} & (i = D)
    \end{dcases}
\end{equation}

\subsection{Mixture Density Networks}

Neural networks are a widely used tool in machine learning due to their ability to learn complex, nonlinear mappings between input variables $\vec{x}$ and output variables $\vec{t}$. Neural networks can model this mapping by learning from a set of training data which provide concrete examples of which output values correspond to each set of input values. Conventionally, neural networks are trained by minimizing the mean squared error between known values from the training data and predicted outputs from the neural network. However, this approach tends to be wholly inadequate for inverse problems, where one set of input values may correspond to multiple output values, or more generally, to some posterior probability density $p(\vec{t} \mid \vec{x})$ (i.e., the probability density of $\vec{t}$ given some input $\vec{x}$). To preserve the practicality of neural networks and extend their functionality to include arbitrary probability functions, \citet{bishop1994MixtureDensity} introduced a class of neural networks called Mixture Density Networks, which combine a conventional neural network with a mixture density model. The posterior $p(\vec{t} \mid \vec{x})$ can be approximated by a linear combination of $m$ kernel functions $\phi_i(\vec{t} \mid \vec{x})$
\begin{equation}
    \label{eq:mixture}
    p(\vec{t} \mid \vec{x}) = \sum_{i=1}^m \alpha_i(\vec{x})\phi_i(\vec{t} \mid \vec{x}),
\end{equation}
where $\alpha_i$ are mixture weights. Various functions can be chosen for $\phi_i(\vec{t} \mid \vec{x})$. We focus here on a mixture model with Gaussian kernels of the form:
\begin{equation}
    \label{eq:gaussian}
    \begin{split}
    \phi_i(\vec{t} \mid \vec{x}) =\ & \frac{1}{(2\pi)^{\nicefrac{c}{2}}\det(\boldsymbol{\Sigma}_i)^{\nicefrac{1}{2}}}\ \times \\&
    \exp\left\{
        -\frac{1}{2}(\vec{t} - \boldsymbol{\mu}_i)^\top \boldsymbol{\Sigma}_i^{-1} (\vec{t} - \boldsymbol{\mu}_i) 
        \right\},
    \end{split}
\end{equation}
where $c$ is the number of kernels in $\vec{t}$ (i.e., the number of output variables), and $\boldsymbol{\mu}_i$ is the center of the $i$th Gaussian kernel with a diagonal covariance matrix $\boldsymbol{\Sigma}_i$:
\begin{equation}
\renewcommand{\arraystretch}{0.5}
    \boldsymbol{\Sigma}_i = 
    \operatorname{diag} (\boldsymbol{\sigma}_i)=
    \begin{bmatrix}
    \sigma_{i,1} &  & \\
     & \ddots & \\
     &  & \sigma_{i,c}
    \end{bmatrix}.
\end{equation}
The conditional probability distribution $p(\vec{t} \mid \vec{x})$ is completely described by weights $\alpha_i$, means $\boldsymbol{\mu}_i$, and variance $\boldsymbol{\sigma}_i$. Training the MDN to predict these outputs therefore allows reconstructing the distribution. With $m$ mixtures and $c$ parameters, the total number of network outputs is $(2c + 1)m$.
A mixture density network is built as a conventional feedforward neural network, where the last layer approximates the distribution parameters. The model can then be trained with a maximum likelihood approach by minimizing the average negative log-likelihood $\mathcal{L}$ across the training data set:
\begin{equation}
\begin{split}
\label{eq:nll}
 \mathcal{L} &= -\frac{1}{N} \sum_{k=1}^N \ln \mathcal{L}_k \\
             &= -\frac{1}{N} \sum_{k=1}^N \ln \left( \sum_{i=1}^m \alpha_i(\vec{x}_k)\phi_i(\vec{t}_k \mid \vec{x}_k) \right),
 \end{split}
\end{equation}
where $N$ is the size of the training data set.

\begin{figure*}[htb!]
    \centering
    \includegraphics[width=\linewidth]{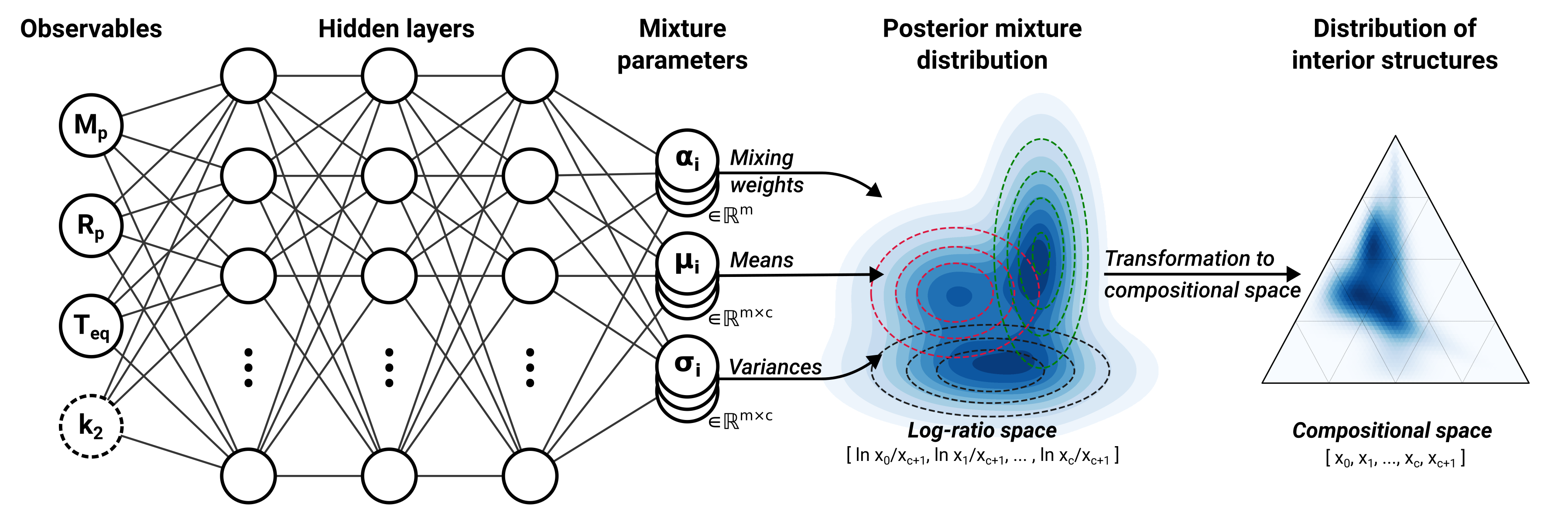}
    \caption{Schematic overview of the MDN architecture and inference procedure.}
    \label{fig:mdn}
\end{figure*}

\subsection{Training data and network architecture}
\label{sec:training_data}

We created a data set of 5.6 million synthetic planets randomly sampled from the prior distributions, summarized in Table~\ref{tab:priors}. The planet mass was chosen uniformly between 0.1 and \SI{25}{\Me}. Each planet was set at a specific equilibrium temperature ranging from 100 to \SI{1000}{K}. The mass fraction of each planetary layer was sampled from the simplex so that they add to one. The gas mass fraction $w\_{Gas}$ was sampled logarithmically with a lower limit of \num{e-8}, while the other mass fractions were sampled uniformly. Given these inputs, the TATOOINE model calculates planet radius and thickness of each layer. For each planet, we also calculated the fluid Love number $k_2$ using the matrix-propagator approach from \citet{padovan2018MatrixpropagatorApproach}. 

To prepare the training data, we log-ratio transformed both mass fractions and radius fractions according to Eq. \eqref{eq:alr} using the core mass and radius as base coordinate ($x_D$ in Eq. \ref{eq:alr}). This is the key difference to our previous work, which enables the prediction of multivariate distributions of mass and radius fractions. The log-ratio transformation enforces the condition that the mass and radius fractions add up to one and allows the network to operate in unbounded real space $\mathbb{R}$ instead of the simplex $\mathcal{S}$, which permits the use of Gaussian kernels as described above.

As a preprocessing step before training and prediction, we also log-transformed the planet mass, which we found to slightly improve the training performance.

We took 70\% of the data set for training, using the remaining 30\% to evaluate the performance of the MDN during training. In addition, we retained a small set of data for final model validation (see Sec. \ref{sec:validation}).

\begin{table}[htb!]
\centering
\caption{Prior distributions of model parameters for training data generation.}
\label{tab:priors}
\begin{tabular}{@{}lll@{}}
\toprule
Parameter & Range & Shape \\ \midrule
$M\_p$ & 0.1 -- 25 $M_\oplus$ & uniform \\
$T\_{eq}$ & 100 -- 1000 K & uniform \\
$w\_{Core}$ & 0 -- 1  & uniform (simplex) \\
$w\_{Mantle}$ & 0 -- 1  & uniform (simplex) \\
$w\_{Water}$ & 0 -- 1 & uniform (simplex) \\
$w\_{Gas}$ & \num{e-8} - 1 & uniform in $\ln w\_{Gas}$ (simplex) \\ \bottomrule
\end{tabular}
\end{table}

We trained the MDN to predict the parameters of the posterior distributions of the log-ratio-transformed mass fractions ($\ln \nicefrac{w\_{Mantle}}{w\_{Core}}$, $\ln \nicefrac{w\_{Water}}{w\_{Core}}$, $\ln \nicefrac{w\_{Gas}}{w\_{Core}}$) and radius fractions ($\ln \nicefrac{d\_{Mantle}}{d\_{Core}}$, $\ln \nicefrac{d\_{Water}}{d\_{Core}}$, $\ln \nicefrac{d\_{Gas}}{d\_{Core}}$). We trained two models with different sets of inputs: Model 1 with $M\_p$, $R\_p$, $T\_{eq}$, and Model 2 with $M\_p$, $R\_p$, $T\_{eq}$, $k_2$.
%

The MDN is built from a feedforward neural network using the Keras framework \citep{chollet2015Keras} and TensorFlow \citep{abadi2015TensorFlowLargeScale}, with the MDN output layer adapted from \citet{martin2019CpmpercussionKerasmdnlayer}.

The best MDN architecture was found through hyperparameter optimization using the KerasTuner framework \citep{omalley2019KerasTuner}. We optimized for the number of hidden layers, the number of units per layer, the learning rate, as well as the batch size. We kept the number of mixture components fixed at $m=50$, because we noticed that the tuner would always optimize for the highest available number of mixtures, but with very small mixture weights for most components. We found 50 components to be a good middle ground where training accuracy was good, but without too many components contributing little to the posterior distribution.

The architecture which yielded the best training performance for Model 1 consists of three hidden layers with 384 nodes per layer with a batch size of 750. For Model 2, the best architecture consists of three hidden layers with 896 nodes per layer with a batch size of 1000. Models with a base learning rate of 0.001 performed best in both cases.

Each hidden layer is activated with a Rectified Linear Unit (ReLU), which is a commonly used activation function in deep learning models \citep{nair2010RectifiedLinear, goodfellow2017DeepLearning}. To ensure that the variances are always positive, we activated $\boldsymbol{\Sigma}_i$ in the output layer with a nonnegative exponential linear unit (NNELU) after \citet{brando2017MixtureDensity}:
\begin{equation}
    \mathrm{NNELU}(x) =    
    \begin{cases}
       x + 1 &\quad \mathrm{for}\:x\geq0\\
       \exp{(x)} &\quad \mathrm{for}\:x<0\\
     \end{cases}
\end{equation}

The nodes for mixture weights $\alpha_i$ and means $\boldsymbol{\mu}_i$ are activated with a linear function to allow unrestricted output values.


To avoid overfitting, we applied an early stopping of the learning algorithm once the validation loss did not improve for 8 consecutive training epochs. 

To improve training performance, we reduced the learning rate by a factor of ten every time the validation loss stops improving for more than four epochs during training, down to a lower bound for the learning rate of \num{e-8}. This helps fine-tune the model weights once a near-optimal set of parameters has been learned.

The MDN was trained on a GPU workstation with eight NVIDIA RTX A5000 graphics cards. The (wall clock) training time for a model was around 3 hours.

\subsection{Backtransformation to mass and radius fractions}
\label{sec:pipeline}
From the predicted parameters of the MDN, the approximate posterior distribution of the log-ratio transformed mass and radius fractions corresponding to the given inputs of the MDN can be reconstructed according to Eq. \eqref{eq:mixture}. The log-ratio space is not particularly useful for interpreting the inferred interior structure distributions. However, the back-transformation of the Gaussian mixture onto the simplex (Eq. \ref{eq:inv_alr}) is mathematically unwieldy, as the normal distributions are highly deformed when in the compositional space. Instead, we randomly sample a sufficiently large number of points from the log-ratio posterior probability distribution and transform these back into compositional space. This is conceptually similar to MCMC sampling and gives a good approximation of the posterior distribution.

\subsection{Incorporating measurement uncertainties}
\label{sec:uncertainty}
The current network architecture is built on the assumption that the input parameters are known exactly without uncertainties. However, except for Solar System planets, observations of exoplanets will always come with considerable measurement uncertainties. With ExoMDN, measurement uncertainties can be taken into account in a straightforward way by repeatedly sampling $n$ times from within the error bars of the input parameters, predict the interior distribution for each sample, and combining the results into a single posterior distribution. This can be either done via summing up each Gaussian mixture in log-ratio space and then normalizing the resulting distribution, or by first subsampling from each prediction $n'$ times and then merging the samples (for a final dataset size of $n\times n'$ samples). Subsampling first and then merging is considerably less memory and processing intensive, as the full posterior distribution can be built up sequentially from each planet sample. Summing up all predicted posterior distributions first requires loading the entire posterior distribution, consisting of $n\times m$ multivariate Gaussian kernels, into memory. Sampling from this mixture distribution can be computationally very expensive for large sample sizes $n$, which are needed to treat the measurement uncertainties well. We find that both approaches bear no functional difference in the predicted full posterior distributions (Fig. \ref{fig:errorsampling_comparison} in the Appendix). We therefore chose the approach of sampling first from each prediction and then merging, as it is {also} easy to implement in the current prediction pipeline. However, $n'$ should be chosen significantly smaller than $n$ to avoid oversampling of specific mass-radius-temperature pairs.

\section{Validation}
\label{sec:validation}

To establish the accuracy of the trained MDN, we validate it in two ways: by forward modeling and by independent inference. In the first case, we used the predicted mass fractions as inputs to the forward model and recomputed the interior structures of planets to investigate how well the planet radius can be retrieved from the predictions. This allows us to put constraints on systematic errors in the MDN outputs. In the second case, we ensured that the MDN predictions are accurate and consistent with other inference methods by comparing the predicted posterior distributions with those obtained by an independent inference approach.

\subsection{Radius accuracy}
\label{sec:forward_error}
\begin{figure*}[ht]
    \centering
    \includegraphics[width=\linewidth]{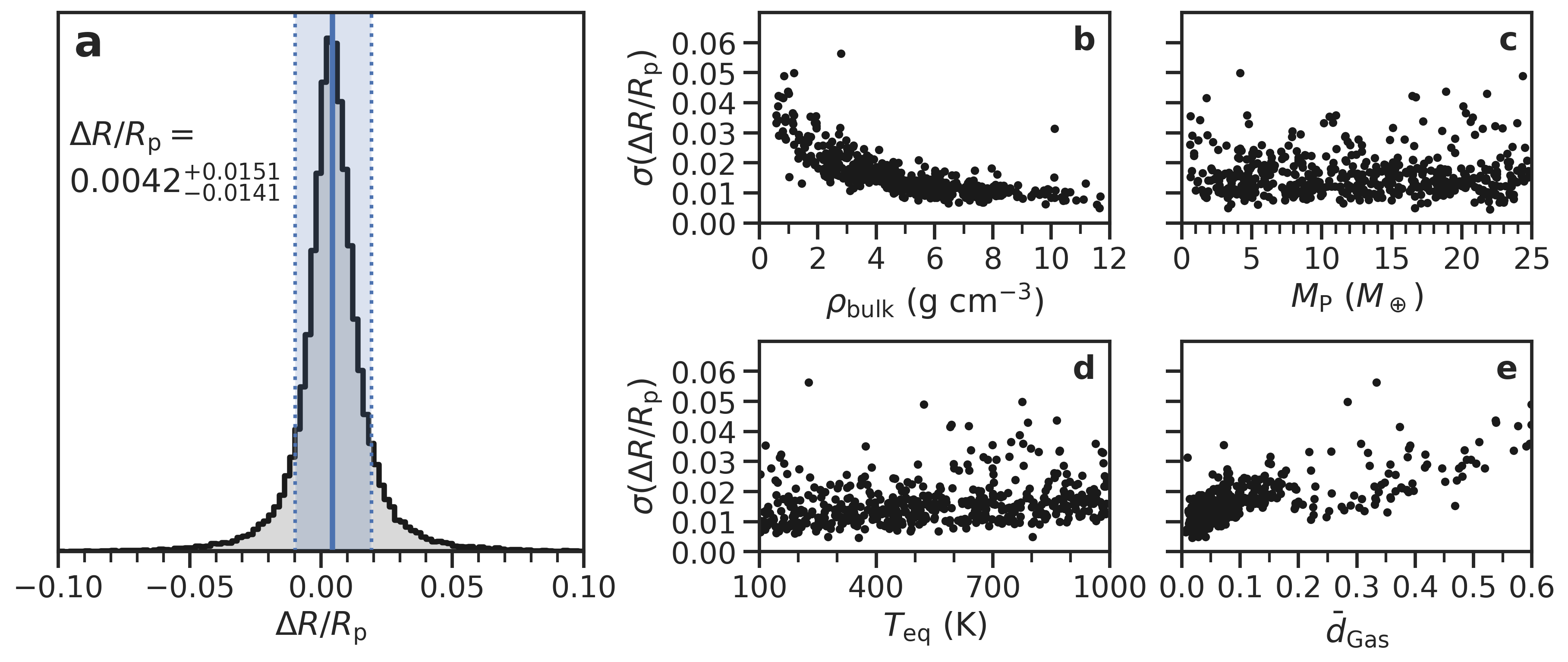}
    \caption{Radius accuracy of Model 1 after recalculating the planet interior based on the MDN prediction. Panel a shows the distribution of the relative radius error of \num{10000} sample points. The blue line marks the median, with the blue area showing the range where 80\% of values lie. Panel b--e show the standard deviation in relative radius errors $\sigma$ for a variety of planet parameters: bulk density (b), planet mass (c), equilibrium temperature (d), average atmosphere thickness $\bar{d}\_{Gas}$ of recomputed planet samples (e). Each point represents one of 500 planets from the test data set (see the text for more details).}
    \label{fig:validation_dr}
\end{figure*}

We used the MDN to predict the interior structures distributions of 500 randomly selected planets out of the test data set. We took 200 samples of interior structures for each prediction and model these planets with the TATOOINE forward model by taking the mass fractions of the layers as inputs (i.e., \num{10000} sample points in total). We then compared the relative error $\frac{\Delta R}{R\_p} = \frac{(R\_{val} - R\_p)}{R\_p}$ between the true planet radius $R\_p$ and the recomputed planet radius $R\_{val}$ obtained from the MDN predictions (Fig.~\ref{fig:validation_dr}).

\begin{figure}[ht]
    \centering
    \includegraphics[width=\linewidth]{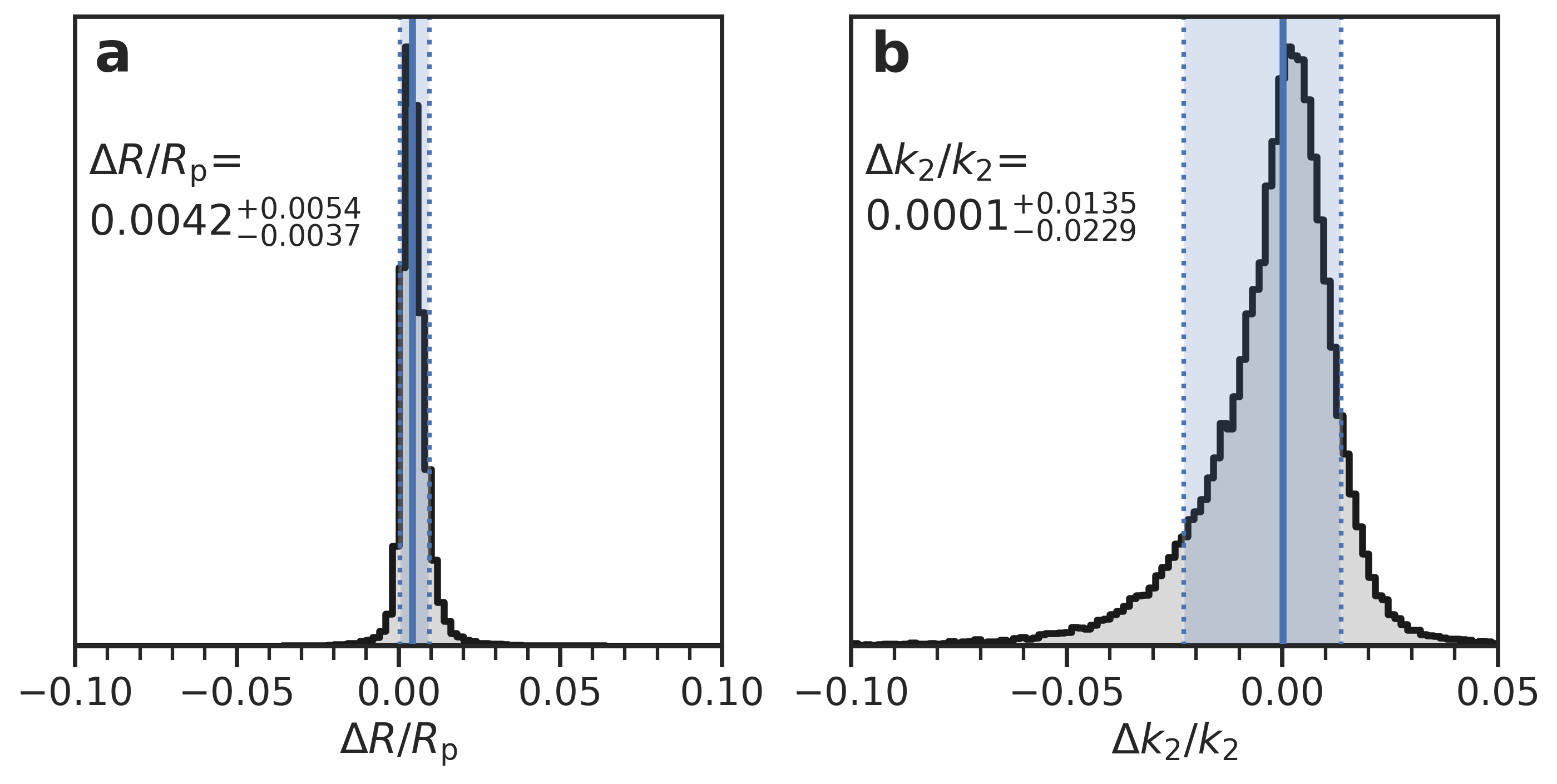}
    \caption{Radius accuracy (a) and $k_2$ accuracy (b) of model 2 after recalculating the planet interior based on the MDN prediction. Panel a shows the distribution in the relative radius error, panel b of the relative $k_2$ error of \num{10000} sample points. Blue lines marks the median, with the blue areas showing the range where 80\% of values lie.}
    \label{fig:validation_drk2}
\end{figure}

We find that the recomputed planet radii of Model 1 fit closely to the expected ones, with the MDN introducing a slight overestimation of the radius of about 0.4\% (Fig.~\ref{fig:validation_dr}a). The MDN introduces a small amount of noise into the recomputed radii, with 80\% of planets having a radius error of less than 1.5\%. The MDN does not perform equally well across the parameter space. Low-density planets tend to have a wider spread in radius errors (Fig.~\ref{fig:validation_dr}b), largely independent of planet mass (Fig.~\ref{fig:validation_dr}c). Higher equilibrium temperatures increase the error slightly (Fig.~\ref{fig:validation_dr}d). We attribute the larger errors mainly to the atmosphere. The recomputed radius errors tend to be the largest in planets with extensive gas envelopes (Fig.~\ref{fig:validation_dr}e). Small errors in the prediction of the gas mass fraction are amplified into larger radius errors due to the low density of the atmosphere. In addition, the transformation from log-ratios to mass and radius fractions amplifies any small uncertainty present in the atmosphere-core log-ratio predictions.

For Model 2, we find that the planetary radii can be reproduced very accurately with a relative radius error of less than 0.55\% (Fig. \ref{fig:validation_drk2}a). As with Model 1, the radius is slightly overestimated by 0.4\%. We additionally check how well the fluid Love number $k_2$ is reproduced by computing the relative error $\frac{\Delta k_2}{k_2} = \frac{(k\_{2,val} - k_2)}{k_2}$, where $k\_{2,val}$ is the Love number of the validation planet to be reproduced, and $k_2$ is the fluid Love number calculated from the predicted interior mass fractions. We find that $k_2$ is reproduced well, with more than 80\% of the points falling within 2.3\% of the true $k_2$ value (Fig. \ref{fig:validation_drk2}b). While the median of the data set sits at zero error, the data set is slightly skewed toward low $k_2$ values. This is most likely caused by the atmosphere. The fluid Love number $k_2$ is highly sensitive to the density structure of the planet, especially in the upper layers. Slight overestimations of the atmosphere mass fractions result in larger underestimations of $k_2$.

\subsection{Independent inference}
\begin{figure}[ht]
    \centering
    \includegraphics[width=\linewidth]{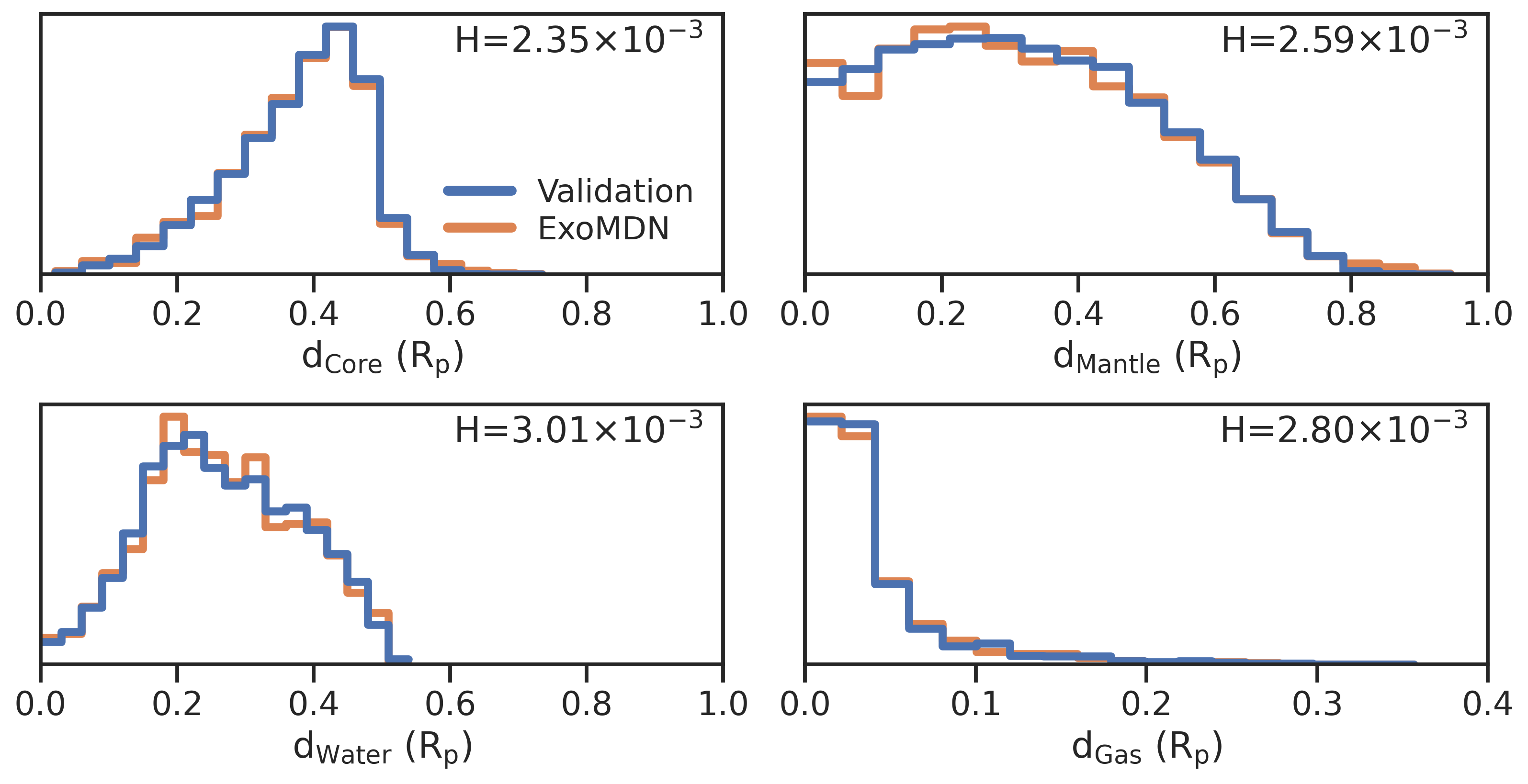}
    \caption{Example of the Hellinger distances for a well-predicted planet (\SI{4.722}{\Me}, \SI{1.82}{\Re}) from the 20 validation planets. The blue line marks the independent validation, the orange line shows the ExoMDN prediction.}
    \label{fig:hellinger_example}
\end{figure}

We randomly selected 20 planets from a test dataset that the MDN did not see during training, and ran an independent inference of their interior structures using a straightforward Monte-Carlo sampling method, assuming a radius uncertainty of 1\%. We assessed how well the predicted posterior distributions $P$ fit to the posterior distributions from the validation set $Q$ by calculating the Hellinger distance $H(P,Q)$ for each marginal distribution following the approach by \citet{haldemann2023ExoplanetCharacterization}. The Hellinger distance is an integrated metric bounded between 0 and 1 that measures the similarity of two probability distributions. Two identical probability distributions have a Hellinger distance of 0, while a Hellinger distance of 1 is reached when there is no overlap between the two distributions. We binned the data into $n=20$ bins with sample frequencies $p_i$ and $q_i$ in each bin. The (squared) Hellinger distance is then given by
\begin{equation}
    H^2(P,Q) = \frac{1}{2} \sum_i^n \left( \sqrt{p_i} - \sqrt{q_i}\right)^2 .
\end{equation}
The average Hellinger distance $\bar{H}$ over the 20 validation planets is shown in Table~\ref{tab:hellinger} for both the log-ratio outputs from the MDN and the transformed compositional mass and radius fractions. We find that the predicted log-ratio distributions compare very well to the validation set, with Hellinger distances around \num{1e-3}. This corresponds to two normal distributions differing in their means by about \num{3e-3} units (assuming a standard deviation of 1), or in the standard deviation by 0.2\% (assuming the same mean). Fig. \ref{fig:hellinger_example} shows an example for a well predicted validation planet with small Hellinger distances.

The transformed mass and radius fractions also fit well, albeit with slightly higher Hellinger distances around \num{3e-3}. The gas mass fraction $w\_{gas}$ is the least well constrained parameter here with $\bar{H}=\num{2.16e-02}$. This mirrors the effect already discussed in Sec. \ref{sec:forward_error}.

\begin{table}[ht!]
\centering
\caption{Average Hellinger distance $\bar{H}$ for 20 randomly selected validation planets of all MDN (log-ratio) output distributions and of their corresponding transformed parameters.}
\label{tab:hellinger}
\begin{tabular}{@{}lc}
\toprule
Parameter                     & $\bar{H}$     \\ \midrule
$\ln d\_{Mantle}/d\_{Core}$ & \num{1.02e-03} \\
$\ln d\_{Water}/d\_{Core}$  & \num{8.84e-04} \\
$\ln d\_{Gas}/d\_{Core}$    & \num{1.30e-03} \\
$\ln w\_{Mantle}/w\_{Core}$ & \num{8.03e-04} \\
$\ln w\_{Water}/w\_{Core}$  & \num{6.87e-04} \\
$\ln w\_{Gas}/w\_{Core}$    & \num{6.07e-04} \\ \midrule
$d\_{Core}$                 & \num{4.35e-03} \\
$d\_{Mantle}$               & \num{2.91e-03} \\
$d\_{Water}$                & \num{3.31e-03} \\
$d\_{Gas}$                  & \num{5.03e-03} \\
$w\_{Core}$                 & \num{2.72e-03} \\
$w\_{Mantle}$               & \num{2.76e-03} \\
$w\_{Water}$                & \num{3.84e-03} \\
$w\_{Gas}$                  & \num{2.16e-02} \\ \bottomrule
\end{tabular}
\end{table}

\section{Results}

\subsection{Earth and Neptune}

We demonstrate the ability of ExoMDN to perform an interior characterization of Earth and Neptune by treating them as if they were exoplanets, where only the mass and radius are measured, and the equilibrium temperature is set according to their orbital distance. Earth represents the archetypical rocky planet whose internal structure is best known of all the planets in the Solar System. Neptune lies on the upper end of the mass range we investigate and is a representative example of volatile-rich planets.

The MDN prediction takes the form of a six-dimensional distribution of the log-ratios of masses and thicknesses of the planetary layers, which can be transformed back to layer mass and thickness as described in Section \ref{sec:pipeline}. For clarity, we will focus in this section only on the thickness of the layers. Figures showing the mass fractions can be found in the Appendix.

Figure~\ref{fig:earth_logrf} shows the posterior distribution of log-ratios for Earth, as approximated by the MDN, given Earth's mass, radius, and equilibrium temperature of \SI{255}{K}. The ellipses in the upper right plots show the location and covariance of each Gaussian kernel, with the colors marking the respective mixture weights $\alpha_i$. The kernels are well spaced with little overlap, and most mixture weights are similar. This indicates that the MDN is able to efficiently leverage all its 50 kernels to construct the posterior distribution.

\begin{figure}[htb!]
    \centering
    \includegraphics[width=\linewidth]{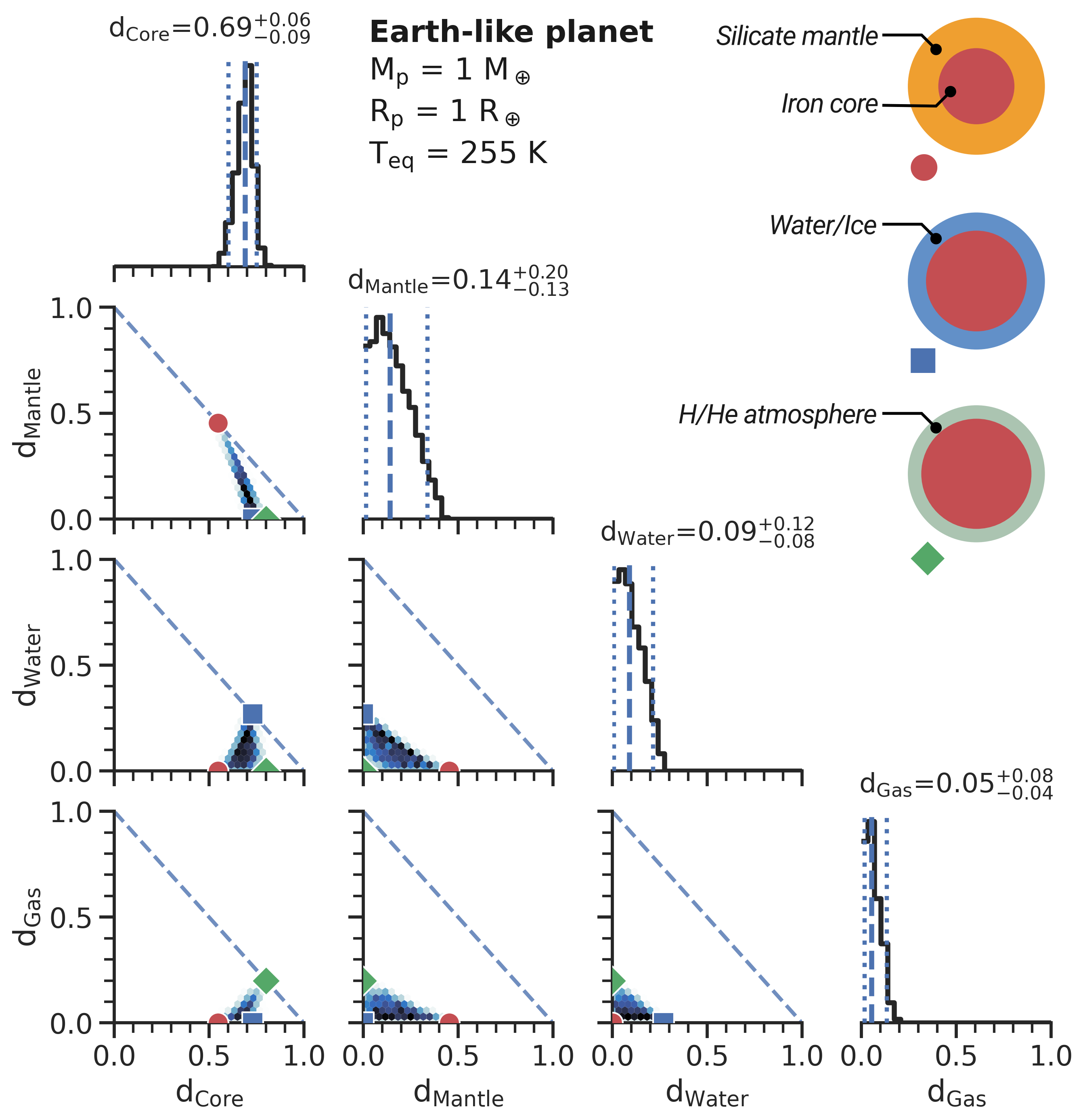}
    \caption{Predicted thickness of interior layers for an Earth-like planet with \SI{1}{\Me} and \SI{1}{\Re}. 
    The colored points mark possible end-member compositions, which are illustrated on the right. The red circle corresponds to Earth's true interior structure. The diagonal plots show the marginal distributions of each layer, with the blue dashed lines marking the median value and the dotted lines the 5th and 95th percentiles.}
    \label{fig:earth_rf}
\end{figure}

We sampled \num{200000} points from the log-ratio distribution to construct the posterior distribution of the actual layer thickness, which are shown in Fig.~\ref{fig:earth_rf}. Given only three observables (mass, radius, equilibrium temperature), the prediction indicates a mostly rocky planet composed of an iron core which makes up at least 50\% of the planet, and with only little water and gas. The colored symbols mark end-member interior structures with only two layers. In this case, only three of these exist, namely: 1) the actual structure of the Earth with an iron core making up 55\% of the interior, and a silicate mantle on top; 2) an iron-water planet composed of a relative core size of 73\% and an ice layer; and 3) an iron-gas planet composed of a massive iron core making up 80\% of the planet, and a H/He envelope taking up the rest.
In cases 2 and 3, the iron core needs to be very large to compensate for the low density of the water and gas layers. Although these two cases are probably not likely to occur in nature, they demonstrate that the interior can not be fully constrained without additional constraints. In this example, the thickness of the silicate mantle in particular can barely be constrained.

It should be noted that due to our assumption of uniform priors, the most commonly predicted interior structures encompass a combination of all four layers. For this reason, the distributions presented should not be understood as definitive probabilities, but rather as the number of potential solutions for each given layer thickness fraction. Consequently, the actual interior of Earth lies outside the bulk of the predicted distribution. In fact, only a single solution exists that matches Earth's mass and radius with only an iron core and a silicate mantle, and no water and extended atmosphere.

\begin{figure}[htb!]
    \centering
    \includegraphics[width=\linewidth]{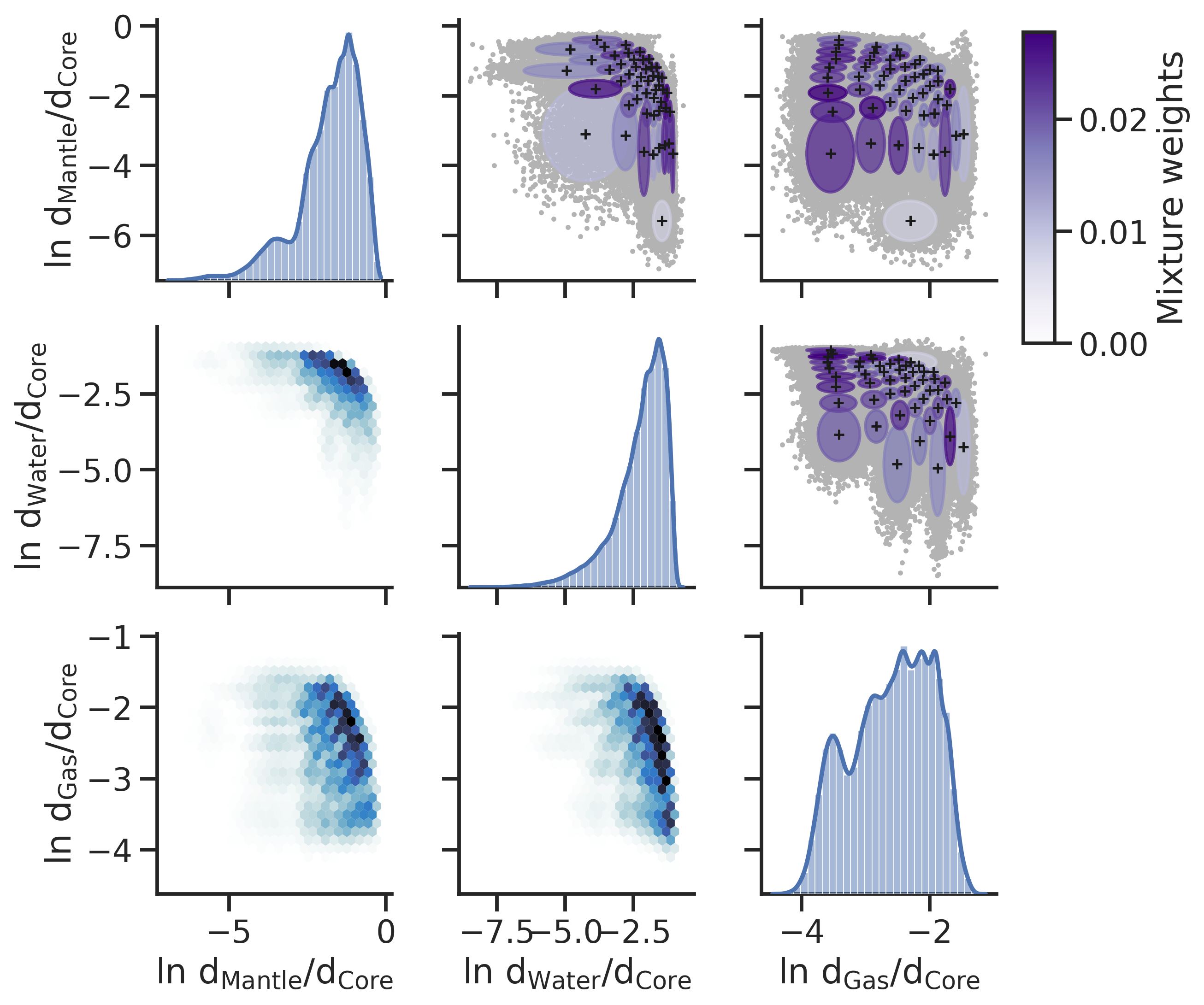}
    \caption{Predicted log-ratios of the thickness of interior layers for an Earth-like planet with \SI{1}{\Me} and \SI{1}{\Re}. The ellipses in the top right plots mark the location and covariance of each of the 50 Gaussian kernels, with the colors showing the mixture weight of each kernel.}
    \label{fig:earth_logrf}
\end{figure}

For Neptune, the MDN predicts a substantial atmosphere between 20 and 70\% of the planet's radius and only a small iron core ($\leq$40\% of the radius). We note here that rather than the actual temperature of \SI{51}{K}, we used an equilibrium temperature of \SI{100}{K}, which is the lowest temperature for which the AQUA EoS used for the water layer is valid. The predicted interior structures lie well within previously published results, which generally agree on Neptune having a small iron-silicate core of about 20\% of the planet's radius and an atmosphere of about 30--40\%, with a water-rich envelope in between \citep[e.g.][]{hubbard1991InteriorStructure, podolak1995ComparativeModels, nettelmann2013NewIndication, neuenschwander2022EmpiricalStructure}. 



\begin{figure}[htb!]
    \centering
    \includegraphics[width=\linewidth]{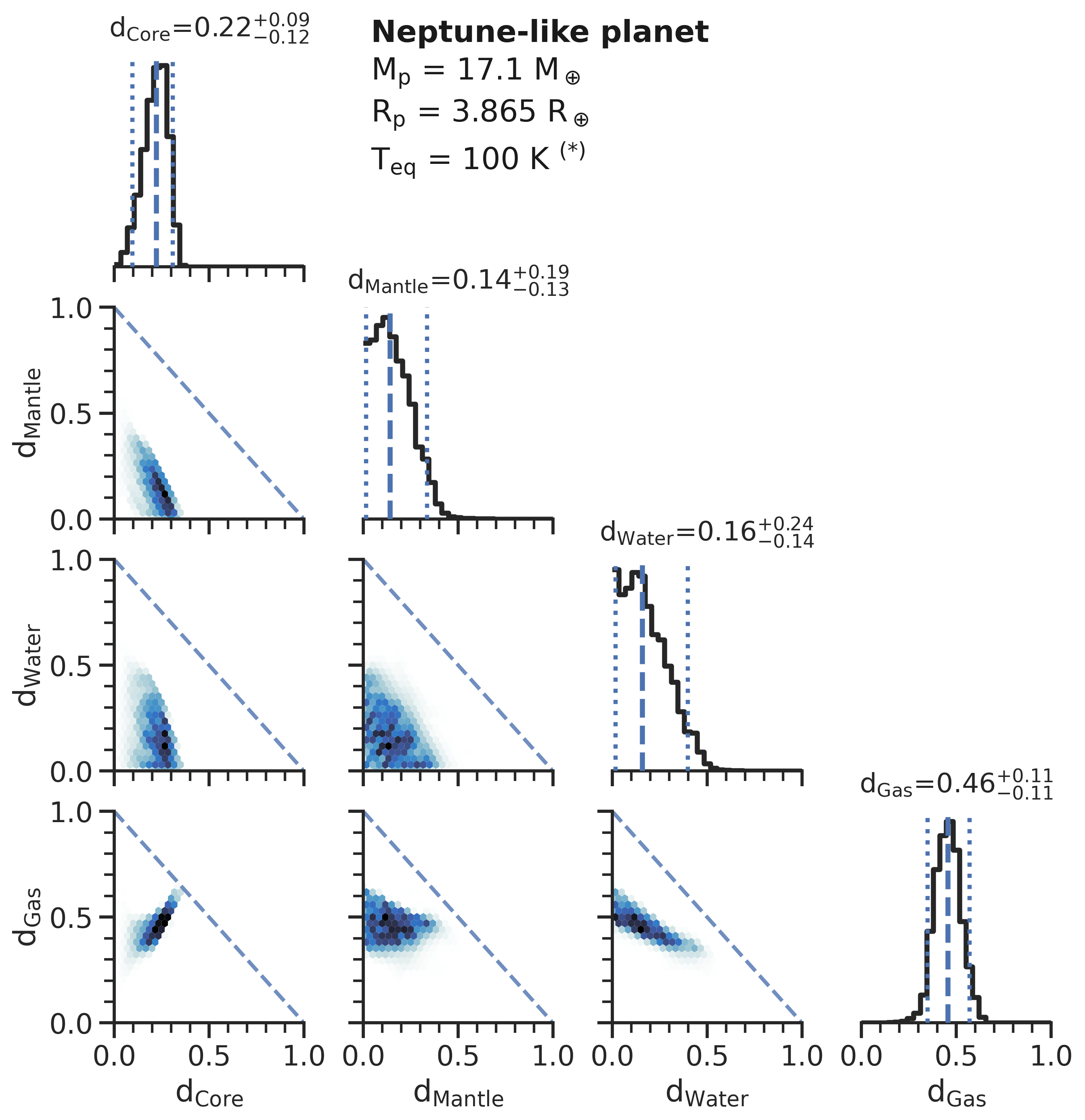}
    \caption{Predicted thickness of interior layers for a Neptune-like planet with \SI{17.1}{\Me} and \SI{3.865}{\Re}. The diagonal plots show the marginal distributions of each layer, with the blue dashed lines marking the median value and the dotted lines the 5th and 95 percentiles.} $^{(*)}$Instead of Neptune's equilibrium temperature of \SI{51}{K}, a value of \SI{100}{K} was used to be in line with the parameter range of the training data.
    \label{fig:neptune_rf}
\end{figure}

\subsection{Application to exoplanets}

One of the main advantages of using neural networks for interior structure inference over other methods such as MCMC sampling is the speed at which the posterior distribution can be obtained. The inference process for MCMC sampling can take several hours per individual planet \citep[e.g.][]{haldemann2023ExoplanetCharacterization}. The MDN can perform the same prediction in fractions of a second. In addition, the MDN model is optimized for bulk processing of inputs owing to the Keras framework, allowing multiple input samples to be predicted simultaneously. Between 1 and 1000 input data points, we find little difference in the computation time needed by the MDN for a prediction ($t\_{MDN}$, Table~\ref{tab:prediction_time}). The sampling from the predicted distribution and transformation to mass and radius fractions (see. Sec. \ref{sec:pipeline}) takes up most of the time ($t\_{sampling}$). Even so, predicting and sampling a thousand planets is possible in under six seconds on a conventional laptop processor\footnote{All predictions were performed on an Intel\textsuperscript{\textregistered} Core\texttrademark{} i5-8250U CPU.}. In fact, the main limitation to predicting a large number of planets simultaneously is the amount of available computer memory.

\begin{table}[htb!]
\centering
\caption{Average inference time for different numbers of planets. $t\_{MDN}$ shows the time needed to retrieve the mixture parameters from the MDN. $t\_{sampling}$ shows the time needed to sample 1000 points from each input planet and transform from log-ratio space to compositional data. For each row, we performed 10 inferences and averaged the computation times.}
\label{tab:prediction_time}
\begin{tabular}{@{}lllll@{}}
\toprule
Planets & $t\_{MDN}$ (s) & $t\_{sampling}$ (s) & Total (s) & \makecell{Time/\\planet (ms)} \\ \midrule
1       & 0.265     & 0.470     & 0.734     & 734.079   \\
10      & 0.257     & 0.524     & 0.781     & 78.126    \\
100     & 0.257     & 0.978     & 1.236     & 12.356    \\
1000    & 0.314     & 4.790     & 5.104     & 5.104     \\ \bottomrule
\end{tabular}
\end{table}

These fast prediction times mean that interior structures can be inferred for every exoplanet for which mass, radius, and equilibrium temperature are known. To demonstrate this, we selected planets from the NASA Exoplanet Archive\footnote{\url{https://exoplanetarchive.ipac.caltech.edu/}} that lie in the parameter space of our training data (Table~\ref{tab:priors}) and for which upper and lower mass and radius uncertainties are given. We used ExoMDN to infer the interior structure of each planet, incorporating the mass, radius, and equilibrium temperature uncertainties according to Sec. \ref{sec:uncertainty}. For each planet, we sampled \num{5000} mass, radius, and equilibrium temperature points from within a normal distribution given by the uncertainties, and predicted the posterior distribution for each point. From each of these posterior distributions, we then generated an additional 10 random samples. In total, this yields \num{50000} samples of interior structures per planet, forming the full posterior distribution and spanning the range of measurement uncertainties.

Table~\ref{tab:exoplanets} in the Appendix shows the 22 planets from this dataset where mass uncertainties are below 10\% and radius uncertainties are below 5\%. This includes the well studied planets GJ 1214 b,  GJ 486 b, and the TRAPPIST-1 planets, among others. For each planet, we provide the predicted median thickness of each interior layer, alongside the ranges in which 90\% of the solutions are found. The total time to produce this data was $\approx \SI{30}{s}$. A more extensive data set of 75 planets with radius and mass uncertainties of 10\% and 20\%, respectively, including both mass fractions and thickness, is available online at the CDS\footnote{via anonymous ftp to cdsarc.cds.unistra.fr (130.79.128.5) or via \url{https://cdsarc.cds.unistra.fr/cgi-bin/qcat?J/A+A/}}.

\begin{figure}[htb!]
    \centering
    \includegraphics[width=\linewidth]{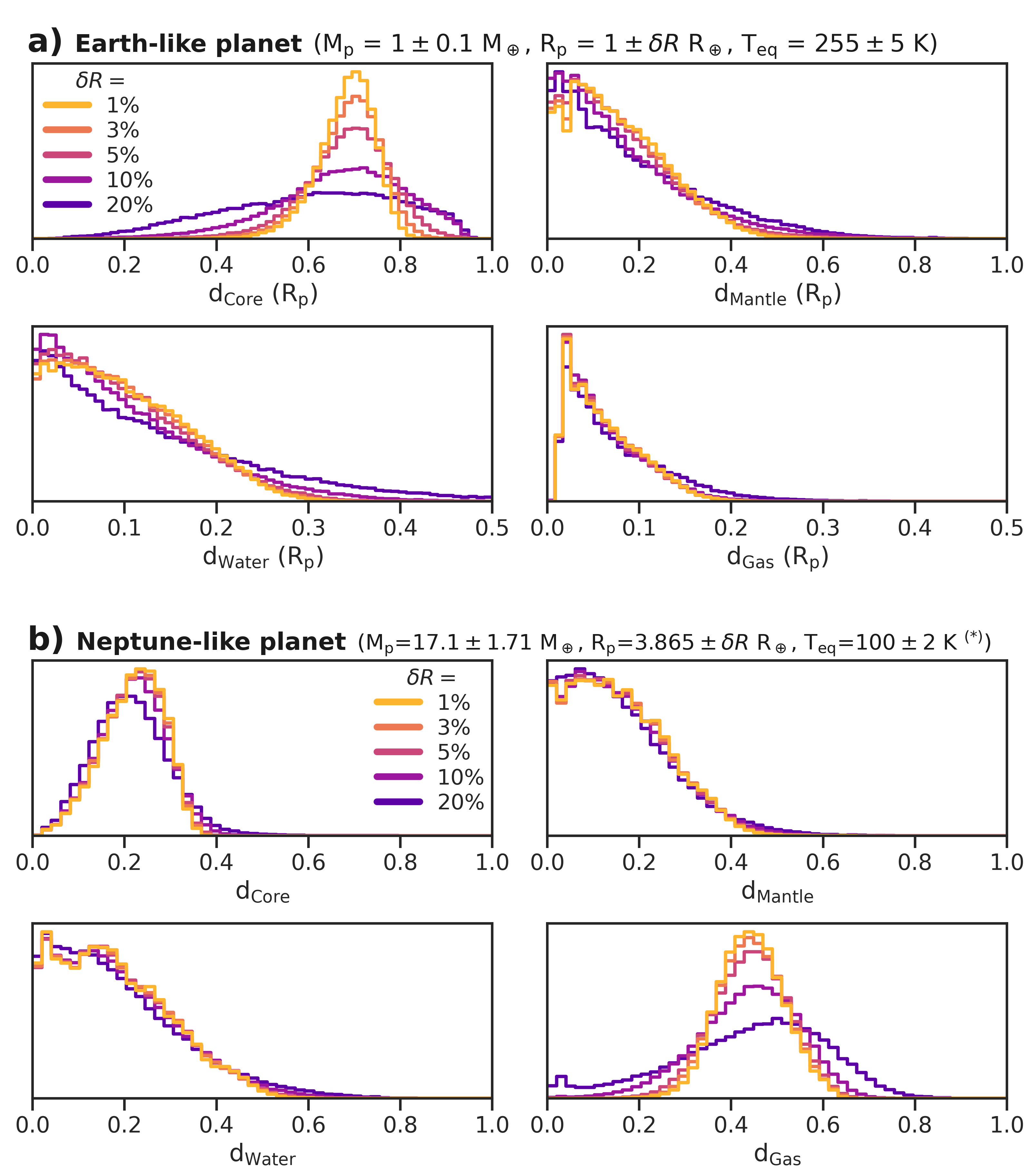}
    \caption{Effect of radius uncertainty $\delta R$ on the ability to constrain the interior for Earth (a) and Neptune (b) analogs. Each panel shows the marginal distributions for each interior layer with increasing amounts of radius uncertainty. An uncertainty of 10\% in mass and 2\% in $T\_{eq}$ has been assumed for both planets in all cases. $^{(*)}$Instead of Neptune's equilibrium temperature of \SI{51}{K}, a value of \SI{100}{K} was used to be in line with the parameter range of the training data.}
    \label{fig:r_err}
\end{figure}

Upcoming exoplanet missions such as PLATO \citep{rauer2014PLATOMission} will significantly increase the number of exoplanets with well-determined masses and radii. PLATO in particular will allow determining the radius of Earth-sized planets with an accuracy of up to $3\%$, while follow-up ground based observations are expected to constrain the mass of these planets with an accuracy of $10\%$ or better. We can leverage the fast prediction times of ExoMDN to investigate the degree to which the interior of a planet could be constrained based on the accuracy of the mass and radius determination. Taking Earth and Neptune as examples, we imposed a $10\%$ mass uncertainty and varied the radius uncertainty between $1\%$ and $20\%$ (Fig.~\ref{fig:r_err}). As above, in each case we sample \num{10000} times from within mass and radius uncertainties, and take 10 random samples from each predicted posterior distribution for a total of \num{100000} samples. 

We find that with a radius accuracy of $3\%$, the core radius fraction of Earth can be constrained to $d\_{Core}=0.69_{-0.14}^{+0.10}$ (error bars are the 5th and 95th percentiles, respectively), which is close to the value we found assuming a perfect knowledge of mass and radius (Fig.~\ref{fig:earth_rf}). With a radius accuracy of $10\%$, the core size is significantly less well constrained ($d\_{Core}=0.68_{-0.27}^{+0.20}$). Similarly, with a radius accuracy of $3\%$, the predicted atmosphere thickness of a Neptune analog (Fig.~\ref{fig:r_err}b) is $d\_{Gas}=0.45_{-0.12}^{+0.12}$, which is again close to the value obtained assuming no error in mass and radius  (Fig.~\ref{fig:neptune_rf}). With a radius accuracy of $10\%$, the uncertainty of $d\_{Gas}$ grows to $d\_{Gas}=0.45_{-0.21}^{+0.16}$. Increasing the radius accuracy has little effect on the possibility to constrain the layers below the atmosphere. Due to the low density of the atmosphere, different planet radii can be easily accommodated by small changes in atmosphere mass without significantly affecting the other layers. In a sense, the presence of a large atmosphere obscures the inference of the thickness of the deeper layers.

The radius accuracy controls to a large extent the uncertainties in the predicted thickness of the various layers. For completeness, we show in  Fig.~\ref{fig:mr_err} in the Appendix predictions of Earth- and Neptune-like interiors obtained when both radius and mass accuracies are varied simultaneously (from 1\% to 20\% and from 3\% to 40\%, respectively). Indeed, the inferred structures are very similar to those we obtained by fixing the mass accuracy to 10\%.

\subsection{Constraining the interior with $k_2$}
Mass, radius, and equilibrium temperature alone are not sufficient to fully constrain the interior of a planet, as demonstrated above. The fluid Love number $k_2$ is a potential direct link from observation to interior structure, as it only depends on the density distribution in the planet. This stands in contrast to for example the elemental abundances of the host star, which may be representative of the bulk abundances of the planet and its atmosphere \citep{dorn2015CanWe, dorn2017BayesianAnalysis, brugger2017ConstraintsSuperEarth, spaargaren2020InfluenceBulk}, but which necessitate additional assumptions about the planet formation and evolution history.

\begin{figure}[htb!]
    \centering
    \includegraphics[width=\linewidth]{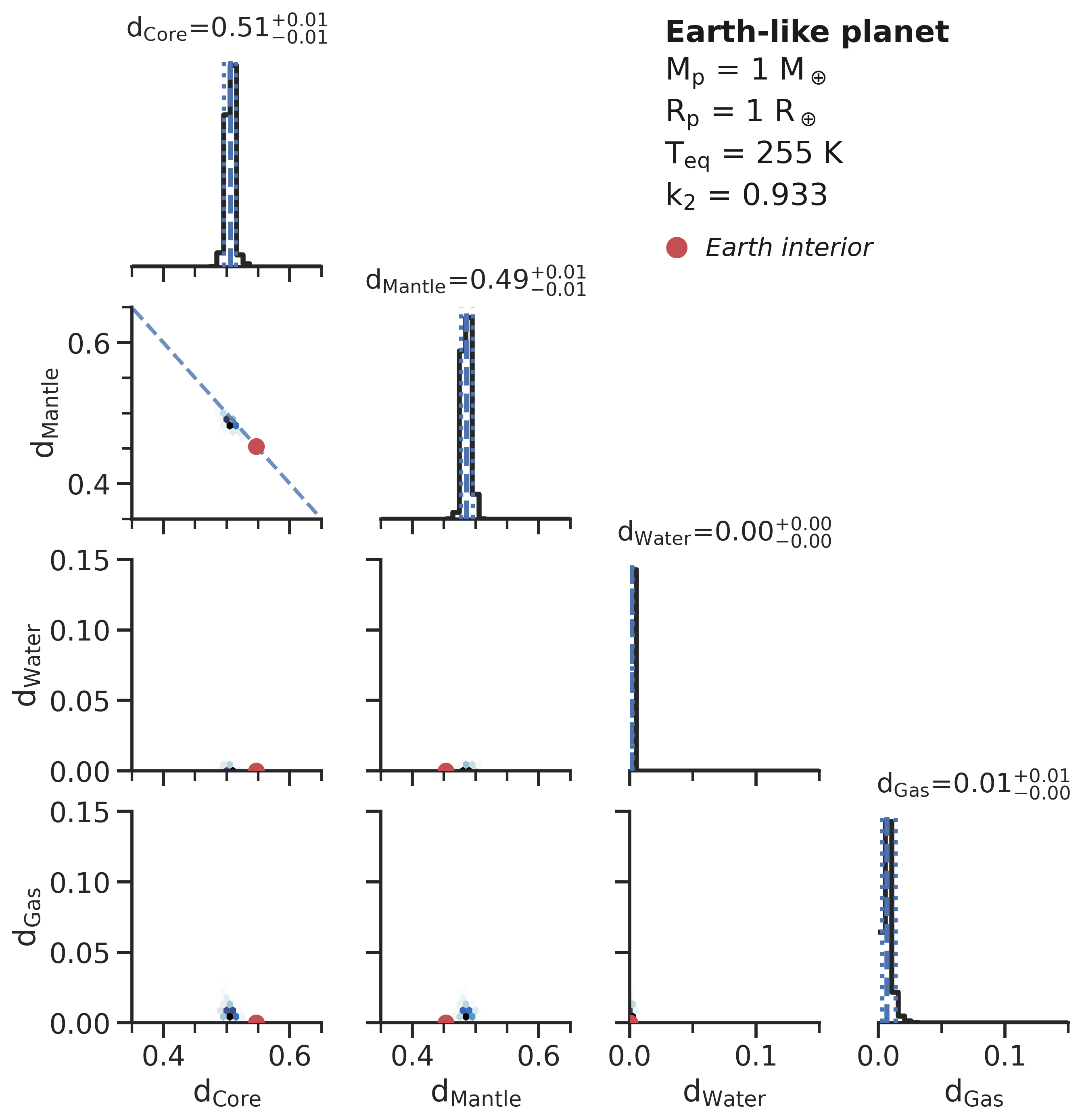}
    \caption{Predicted thickness of interior layers for an Earth-like planet with \SI{1}{\Me} and \SI{1}{\Re}, where also $k_2$ is known ($k_2=0.933$) in addition to mass and radius. The red circle marks Earth's true interior structure. The diagonal plots show the marginal distributions of each layer, with the blue dashed lines marking the median value and the dotted lines the 5th and 95th percentiles. Compared to Fig.~\ref{fig:earth_rf} the axis limits have been adjusted to better show the model results.}
    \label{fig:earth_rf_k2}
\end{figure}

\begin{figure}[htb!]
    \centering
    \includegraphics[width=\linewidth]{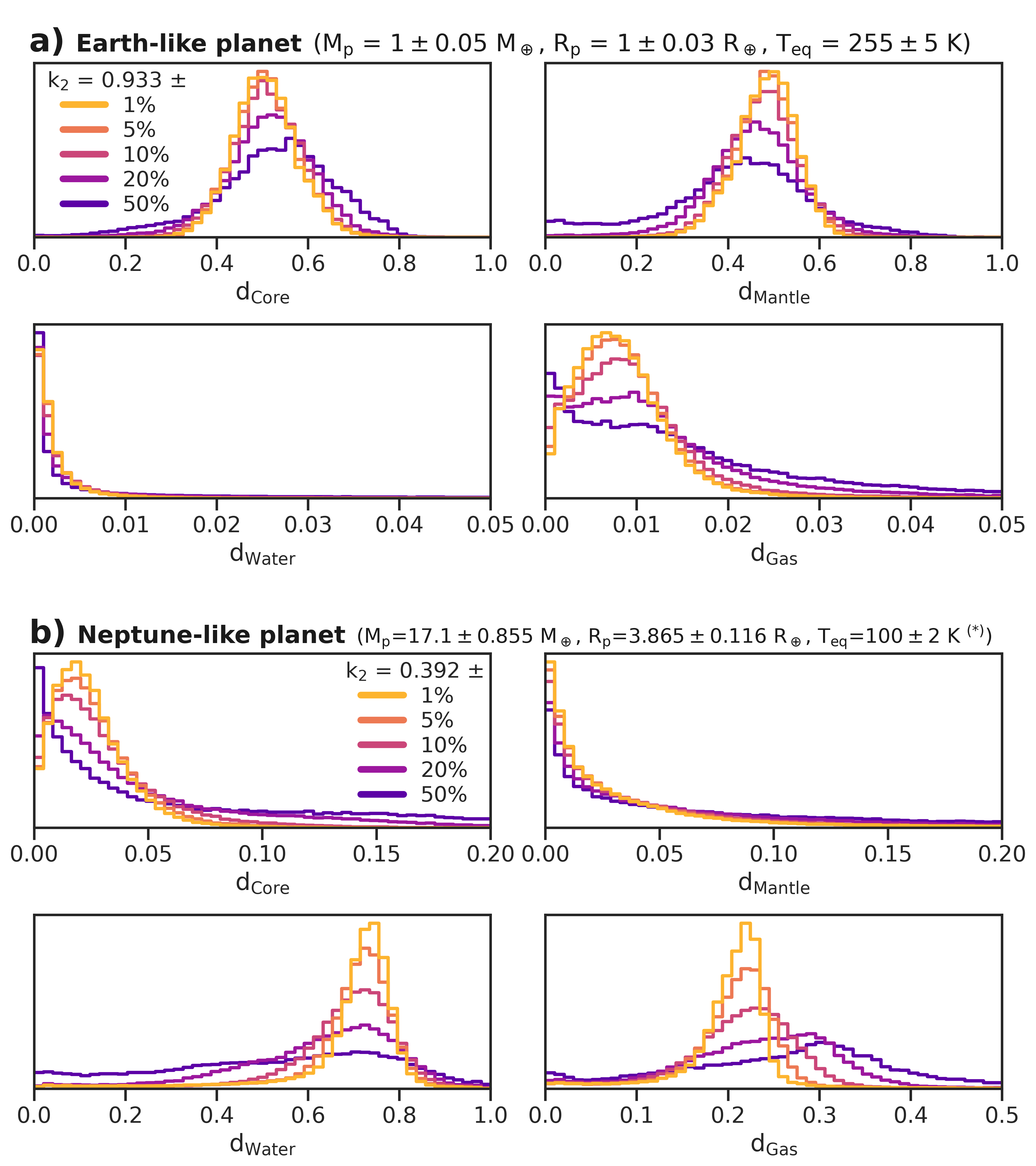}
    \caption{Effect of $k_2$ uncertainty on the ability to constrain the interior for Earth (a) and Neptune (b) analogs. Each panel shows the marginal distributions for each interior layer with increasing amounts of $k_2$ uncertainty. An uncertainty of 5\% in mass, 3\% in radius, and 2\% in $T\_{eq}$ has been assumed for both planets in all cases. $^{(*)}$Instead of Neptune's equilibrium temperature of \SI{51}{K}, a value of \SI{100}{K} was used to be in line with the parameter range of the training data.}
    \label{fig:k2err}
\end{figure}

Figure~\ref{fig:earth_rf_k2} shows the MDN prediction of the interior of Earth, given knowledge of Earth's value of $k_2=0.933$ \citep{lambeck1980EarthVariable}. With this added information, the MDN is capable of fully constraining Earth's actual interior (particular in comparison to Figure~\ref{fig:earth_rf}). In fact, the constraints from $k_2$ are strong enough that the composition of the iron core becomes important. The planets in the training data are modeled with a pure iron core, while Earth has about 10-15 wt.\% of lighter elements in its core \citep{poirier1994LightElements}. Thus, the MDN predicts a smaller core radius of 51\%, while the true core size is about 54.5\% of the total radius. In practice, of course, the measurements $k_2$ for exoplanets will be associated with considerable uncertainties. Constraining a planet's interior to the degree shown in Figure~\ref{fig:earth_rf_k2} is therefore unlikely in the near future. Nevertheless, we can utilize the fast predictions of the MDN to estimate the accuracy that would be needed to properly constrain the interior. We performed a number of predictions for Earth and Neptune analogs with increasing  $k_2$ uncertainties, in addition to mass, radius, and equilibrium temperature uncertainties of 5\%, 3\%, and 2\%, respectively). These values are representative of a very well studied and characterized exoplanet, which would likely be needed for an accurate measurement of the Love number. For Neptune, we take a value of $k_2=0.392$, which we calculated after \citet{hubbard1984PlanetaryInteriors} from the gravitational moment $J_2=\num{3.40843e-3}$ \citep{jacobson2009ORBITSNEPTUNIAN}. The predicted results are detailed in Figure~\ref{fig:k2err}. As the uncertainty in $k_2$ grows, the interior of both planets becomes less and less constrained. We find that with a $k_2$ uncertainty of $10\%$, Earth's core and mantle thickness could be constrained to about $\pm13\%$ of their actual value (within the 5\% and 95\% percentiles). With a $k_2$ uncertainty of $20\%$, mantle and core can be constrained to within $\pm17\%$. Even with large $k_2$ uncertainties, Earth could be clearly identified as a rocky planet with very little water and a thin atmosphere. The uncertainties of mass and radius put a limit on how well the interior can be determined. With the given mass and radius uncertainties, we find that, in the Earth-like case, $k_2$ uncertainties lower than 10\% do not constrain the interior further. For Neptune, a $10\%$ uncertainty in $k_2$ could help constrain the atmospheric thickness to $22_{-13}^{+10}\%$ of the planet's radius.

\section{Discussion and Conclusions}

MDNs can provide a reliable way to rapidly characterize the interior structure of exoplanets within fractions of a second. Compositional data such as mass fractions of individual interior layers can be easily accommodated in the network by using log-ratio transformations. An additional benefit of a machine learning approach over other inference methods is that the forward model computations are decoupled from the actual inference process. The training data is calculated separately before training, and the training process encodes the information from the training data into the network weights. The trained network itself is stand-alone, and interior inferences can be performed without requiring the training data, a dedicated interior model, a separate inference scheme, or prior expertise about exoplanet interior modeling. This stands in contrast to MCMC sampling, where running the data-generating forward model during the inference is an integral part of exploring the posterior distributions. 

While in this work the training data was generated from a single forward model, this is actually not necessary for the training of the network. Since the forward model data generation is separated from the inference itself, different parts of the data set can be modeled by different dedicated forward models, for example to include both Jupiter-like and low-mass planets which may require different modeling approaches. Importantly, this means that the training data can be computed, collected, and combined from multiple sources without much overhead and without the need to integrate different numerical codes into a single model, as would be needed for MCMC sampling. Furthermore, this means that this method is easily extendable to different models and applicable to other inverse problems. However, the necessary prior generation of training data locks the model assumptions of the forward model into the training data. Changing the forward model therefore requires computing a new set of training data and training of a new neural network. This may be a drawback if the forward model assumptions often change (e.g., with different atmosphere compositions).

Our method is best suited for problems where the number of constraining parameters is relatively small. The required number of training samples increases (potentially exponentially) with each additional parameter, a phenomenon which has been termed the \enquote{curse of dimensionality} \citep{bellman1966DynamicProgramming}. This can make the generation and handling of training data cumbersome and time-consuming for larger numbers of constraining parameters, and generally one needs to sample from the entire investigated parameter space to achieve good MDN performance. A potential way to alleviate this issue could be to generate the training data \enquote{on the fly} and train the network with an incremental learning approach \citep{vandeven2022ThreeTypes}, where the network learns continuously with new incoming data, thus reducing the need to save large amounts of data.

Conditional invertible neural networks (cINN) may be another alternative, as demonstrated by \citet{haldemann2023ExoplanetCharacterization}. These potentially work better with higher-dimensional data while requiring comparatively less training data, with the tradeoff that the network setup is more complex and predictions are generally slower.

As with other machine learning methods, the nature of the training process introduces a small amount of intrinsic noise into the model. However, we have shown that the errors introduced by this are generally small (see Section \ref{sec:forward_error}), particularly for exoplanets where uncertainties in the observable quantities are relatively large.

The file size of the fully trained model is only $\approx \SI{6.8}{MB}$, which facilitates sharing and distribution online. The posterior distributions predicted by ExoMDN can provide a first characterization of newly observed planets, which can then be further explored with dedicated models. ExoMDNs posterior distributions could also be employed as advanced priors for MCMC inferences based on more sophisticated forward models to help speed up their convergence. We believe that ExoMDN is a valuable tool for the exoplanet science community to provide a rapid first characterization of the possible interiors of low-mass planets.


\begin{acknowledgements}
We thank Heike Rauer for important suggestions and discussions on the role of uncertainties in mass and radius, and an anonymous referee for their comments, which helped improve a previous version of the manuscript. We acknowledge the support of the DFG priority program SPP 1992 \enquote{Exploring the Diversity of Extrasolar Planets} (TO 704/3-1) and of the research unit FOR 2440 \enquote{Matter under planetary interior conditions} (PA  3689/1-1). Training of ExoMDN was performed on a GPU Workstation also sponsored by the DFG research unit FOR 2440 (grant number RE 882/19-2), which is gratefully acknowledged.  
\end{acknowledgements}

\bibliographystyle{aa}
\bibliography{manuscript.bib}{}



\begin{appendix}
\onecolumn
\section{Exoplanet predictions}

\begin{table*}[ht!]
\rotatebox{90}{
\resizebox{0.9\textheight}{!}{
\begin{minipage}{0.9\textheight}
\centering
\caption{Predicted thicknesses $d_i$ of interior layers for exoplanets with well known masses and radii.}
\label{tab:exoplanets}
\begin{tabular}{@{}lllllllll@{}}
\toprule

Planet & Observation &  & & & Prediction &  &  &  \\
& Mass (\si{\Me}) & Radius (\si{\Re}) & $T\_{eq}$ (K) & References & $d\_{Core}$ & $d\_{Mantle}$ & $d\_{Water}$ & $d\_{Gas}$ \\ \midrule

GJ 1214 b & 8.17$\pm$0.43 & 2.742$\pm$0.053 & 596.0$\pm$19.0 & \makecell{Cloutier et al. 2021\\(Carter et al. 2011)} & $\num{0.238}_{\num{-0.137}}^{+\num{0.138}}$ & $\num{0.172}_{\num{-0.153}}^{+\num{0.245}}$ & $\num{0.387}_{\num{-0.32}}^{+\num{0.251}}$ & $\num{0.183}_{\num{-0.0881}}^{+\num{0.182}}$ \\
GJ 486 b & 2.82$\pm$0.12 & 1.305$\pm$0.067 & 701.0$\pm$13.0 & Trifonov et al. 2021 & $\num{0.74}_{\num{-0.124}}^{+\num{0.106}}$ & $\num{0.0812}_{\num{-0.0736}}^{+\num{0.169}}$ & $\num{0.0767}_{\num{-0.0586}}^{+\num{0.109}}$ & $\num{0.0686}_{\num{-0.0328}}^{+\num{0.0849}}$ \\
HD 136352 b & 4.72$\pm$0.42 & 1.664$\pm$0.043 & 905.0$\pm$14.0 & \makecell{Delrez et al. 2021\\(Kane et al. 2020)} & $\num{0.614}_{\num{-0.125}}^{+\num{0.0827}}$ & $\num{0.143}_{\num{-0.127}}^{+\num{0.228}}$ & $\num{0.124}_{\num{-0.0966}}^{+\num{0.14}}$ & $\num{0.0902}_{\num{-0.0525}}^{+\num{0.108}}$ \\
HD 136352 c & 11.24$\pm$0.65 & 2.916$\pm$0.075 & 677.0$\pm$11.0 &  \makecell{Delrez et al. 2021\\(Kane et al. 2020)} & $\num{0.231}_{\num{-0.133}}^{+\num{0.14}}$ & $\num{0.176}_{\num{-0.156}}^{+\num{0.247}}$ & $\num{0.424}_{\num{-0.354}}^{+\num{0.243}}$ & $\num{0.153}_{\num{-0.0996}}^{+\num{0.204}}$ \\
HD 191939 b & 10.4$\pm$0.9 & 3.39$\pm$0.07 & 893.0$\pm$36.0 & Lubin et al. 2022 & $\num{0.216}_{\num{-0.125}}^{+\num{0.115}}$ & $\num{0.152}_{\num{-0.136}}^{+\num{0.217}}$ & $\num{0.267}_{\num{-0.231}}^{+\num{0.291}}$ & $\num{0.338}_{\num{-0.133}}^{+\num{0.151}}$ \\
HD 219134 c & 4.36$\pm$0.22 & 1.511$\pm$0.047 & 782.0$\pm$6.0 & Gillon et al. 2017 & $\num{0.695}_{\num{-0.0989}}^{+\num{0.0771}}$ & $\num{0.104}_{\num{-0.0934}}^{+\num{0.176}}$ & $\num{0.0962}_{\num{-0.0758}}^{+\num{0.113}}$ & $\num{0.0781}_{\num{-0.0421}}^{+\num{0.088}}$ \\
HIP 97166 b & 20.0$\pm$1.5 & 2.74$\pm$0.13 & 757.0$\pm$25.0 & MacDougall et al. 2021 & $\num{0.356}_{\num{-0.179}}^{+\num{0.135}}$ & $\num{0.267}_{\num{-0.239}}^{+\num{0.326}}$ & $\num{0.294}_{\num{-0.22}}^{+\num{0.208}}$ & $\num{0.0633}_{\num{-0.0435}}^{+\num{0.128}}$ \\
Kepler-36 b & 3.83$\pm$0.11 & 1.498$\pm$0.061 & 978.0$\pm$11.0 & \makecell{Vissapragada et al. 2020\\(Carter et al. 2012)} & $\num{0.68}_{\num{-0.115}}^{+\num{0.0899}}$ & $\num{0.103}_{\num{-0.0926}}^{+\num{0.19}}$ & $\num{0.101}_{\num{-0.0763}}^{+\num{0.117}}$ & $\num{0.0846}_{\num{-0.0425}}^{+\num{0.101}}$ \\
Kepler-36 c & 7.13$\pm$0.18 & 3.679$\pm$0.096 & 928.0$\pm$10.0 & \makecell{Vissapragada et al. 2020\\(Carter et al. 2012)} & $\num{0.193}_{\num{-0.109}}^{+\num{0.0942}}$ & $\num{0.13}_{\num{-0.116}}^{+\num{0.176}}$ & $\num{0.192}_{\num{-0.167}}^{+\num{0.242}}$ & $\num{0.463}_{\num{-0.114}}^{+\num{0.12}}$ \\
LHS 1140 b & 6.38$\pm$0.46 & 1.635$\pm$0.046 & 379.0$\pm$4.0 & \makecell{Lillo-Box et al. 2020\\(Dittmann et al. 2017)} & $\num{0.659}_{\num{-0.102}}^{+\num{0.0858}}$ & $\num{0.159}_{\num{-0.139}}^{+\num{0.199}}$ & $\num{0.121}_{\num{-0.105}}^{+\num{0.129}}$ & $\num{0.0445}_{\num{-0.0282}}^{+\num{0.0596}}$ \\
LHS 1140 c & 1.76$\pm$0.17 & 1.169$\pm$0.038 & 709.0$\pm$8.0 & Lillo-Box et al. 2020 & $\num{0.729}_{\num{-0.113}}^{+\num{0.0784}}$ & $\num{0.0861}_{\num{-0.078}}^{+\num{0.16}}$ & $\num{0.0877}_{\num{-0.0605}}^{+\num{0.0996}}$ & $\num{0.0699}_{\num{-0.0327}}^{+\num{0.0901}}$ \\
LHS 1478 b & 2.33$\pm$0.2 & 1.242$\pm$0.051 & 595.0$\pm$10.0 & Soto et al. 2021 & $\num{0.73}_{\num{-0.124}}^{+\num{0.0945}}$ & $\num{0.0933}_{\num{-0.084}}^{+\num{0.177}}$ & $\num{0.078}_{\num{-0.0641}}^{+\num{0.115}}$ & $\num{0.0689}_{\num{-0.037}}^{+\num{0.0847}}$ \\
TOI-220 b & 13.8$\pm$1.0 & 3.03$\pm$0.15 & 805.0$\pm$21.0 & Hoyer et al. 2021 & $\num{0.243}_{\num{-0.136}}^{+\num{0.134}}$ & $\num{0.183}_{\num{-0.162}}^{+\num{0.254}}$ & $\num{0.407}_{\num{-0.334}}^{+\num{0.236}}$ & $\num{0.145}_{\num{-0.104}}^{+\num{0.225}}$ \\
TOI-270 c & 6.15$\pm$0.37 & 2.355$\pm$0.064 & 488.0$\pm$12.0 & Van Eylen et al. 2021 & $\num{0.27}_{\num{-0.148}}^{+\num{0.137}}$ & $\num{0.201}_{\num{-0.178}}^{+\num{0.264}}$ & $\num{0.394}_{\num{-0.306}}^{+\num{0.237}}$ & $\num{0.121}_{\num{-0.0873}}^{+\num{0.167}}$ \\
TOI-270 d & 4.78$\pm$0.43 & 2.133$\pm$0.058 & 387.0$\pm$10.0 & Van Eylen et al. 2021 & $\num{0.291}_{\num{-0.158}}^{+\num{0.141}}$ & $\num{0.215}_{\num{-0.19}}^{+\num{0.279}}$ & $\num{0.376}_{\num{-0.285}}^{+\num{0.232}}$ & $\num{0.103}_{\num{-0.0788}}^{+\num{0.16}}$ \\
TOI-421 c & 16.42$\pm$1.06 & 5.09$\pm$0.16 & 674.0$\pm$11.0 & Carleo et al. 2020 & $\num{0.165}_{\num{-0.0927}}^{+\num{0.0709}}$ & $\num{0.106}_{\num{-0.0946}}^{+\num{0.145}}$ & $\num{0.122}_{\num{-0.109}}^{+\num{0.171}}$ & $\num{0.593}_{\num{-0.0928}}^{+\num{0.0873}}$ \\
TRAPPIST-1 b & 1.374$\pm$0.069 & 1.116$\pm$0.014 & 400.0$\pm$8.0 & \makecell{Agol et al. 2021\\(Gillon et al. 2017)} & $\num{0.675}_{\num{-0.1}}^{+\num{0.0712}}$ & $\num{0.135}_{\num{-0.12}}^{+\num{0.203}}$ & $\num{0.0965}_{\num{-0.0831}}^{+\num{0.129}}$ & $\num{0.0662}_{\num{-0.0427}}^{+\num{0.0918}}$ \\
TRAPPIST-1 c & 1.308$\pm$0.056 & 1.097$\pm$0.014 & 342.0$\pm$7.0 & \makecell{Agol et al. 2021\\(Gillon et al. 2017)} & $\num{0.674}_{\num{-0.1}}^{+\num{0.0711}}$ & $\num{0.14}_{\num{-0.125}}^{+\num{0.207}}$ & $\num{0.0967}_{\num{-0.0858}}^{+\num{0.131}}$ & $\num{0.0619}_{\num{-0.0408}}^{+\num{0.0879}}$ \\
TRAPPIST-1 d & 0.388$\pm$0.012 & 0.788$\pm$0.011 & 288.0$\pm$6.0 & \makecell{Agol et al. 2021\\(Gillon et al. 2017)} & $\num{0.659}_{\num{-0.127}}^{+\num{0.0745}}$ & $\num{0.152}_{\num{-0.135}}^{+\num{0.242}}$ & $\num{0.0917}_{\num{-0.0796}}^{+\num{0.141}}$ & $\num{0.0594}_{\num{-0.0384}}^{+\num{0.118}}$ \\
TRAPPIST-1 e & 0.692$\pm$0.022 & 0.92$\pm$0.013 & 251.0$\pm$5.0 & \makecell{Agol et al. 2021\\(Grimm et al. 2018)} & $\num{0.666}_{\num{-0.113}}^{+\num{0.0744}}$ & $\num{0.153}_{\num{-0.137}}^{+\num{0.224}}$ & $\num{0.0937}_{\num{-0.0835}}^{+\num{0.14}}$ & $\num{0.0571}_{\num{-0.0398}}^{+\num{0.0931}}$ \\
TRAPPIST-1 f & 1.039$\pm$0.031 & 1.045$\pm$0.013 & 219.0$\pm$4.0 & Agol et al. 2021 & $\num{0.64}_{\num{-0.117}}^{+\num{0.0774}}$ & $\num{0.175}_{\num{-0.155}}^{+\num{0.235}}$ & $\num{0.111}_{\num{-0.0988}}^{+\num{0.143}}$ & $\num{0.0511}_{\num{-0.0375}}^{+\num{0.0824}}$ \\
TRAPPIST-1 g & 1.321$\pm$0.038 & 1.129$\pm$0.015 & 199.0$\pm$4.0 & \makecell{Agol et al. 2021\\(Gillon et al. 2017)} & $\num{0.617}_{\num{-0.123}}^{+\num{0.0807}}$ & $\num{0.193}_{\num{-0.17}}^{+\num{0.244}}$ & $\num{0.122}_{\num{-0.109}}^{+\num{0.147}}$ & $\num{0.0486}_{\num{-0.0366}}^{+\num{0.0766}}$ \\
\midrule
\multicolumn{9}{p{\textwidth}}{Notes: Given are the median of each predicted distribution and the 5\% and 95\% percentiles. For planets with nonsymmetric mass or radius uncertainties, the larger value was used. Separate references for $T\_{eq}$ are given in parentheses. For planets where no uncertainty of $T\_{eq}$ was given, an uncertainty of 2\% is assumed. A more extensive data set of 75 planets with radius and mass uncertainties of 10\% and 20\%, respectively, including both mass fractions and thickness, is available in electronic form at the CDS via anonymous ftp to cdsarc.cds.unistra.fr (130.79.128.5) or via \url{https://cdsarc.cds.unistra.fr/cgi-bin/qcat?J/A+A/}}
\end{tabular}
\end{minipage}
}
}
\end{table*}
\FloatBarrier

\twocolumn

\section{Additional figures}

\begin{figure}[htb!]
    \centering
    \includegraphics[width=\linewidth]{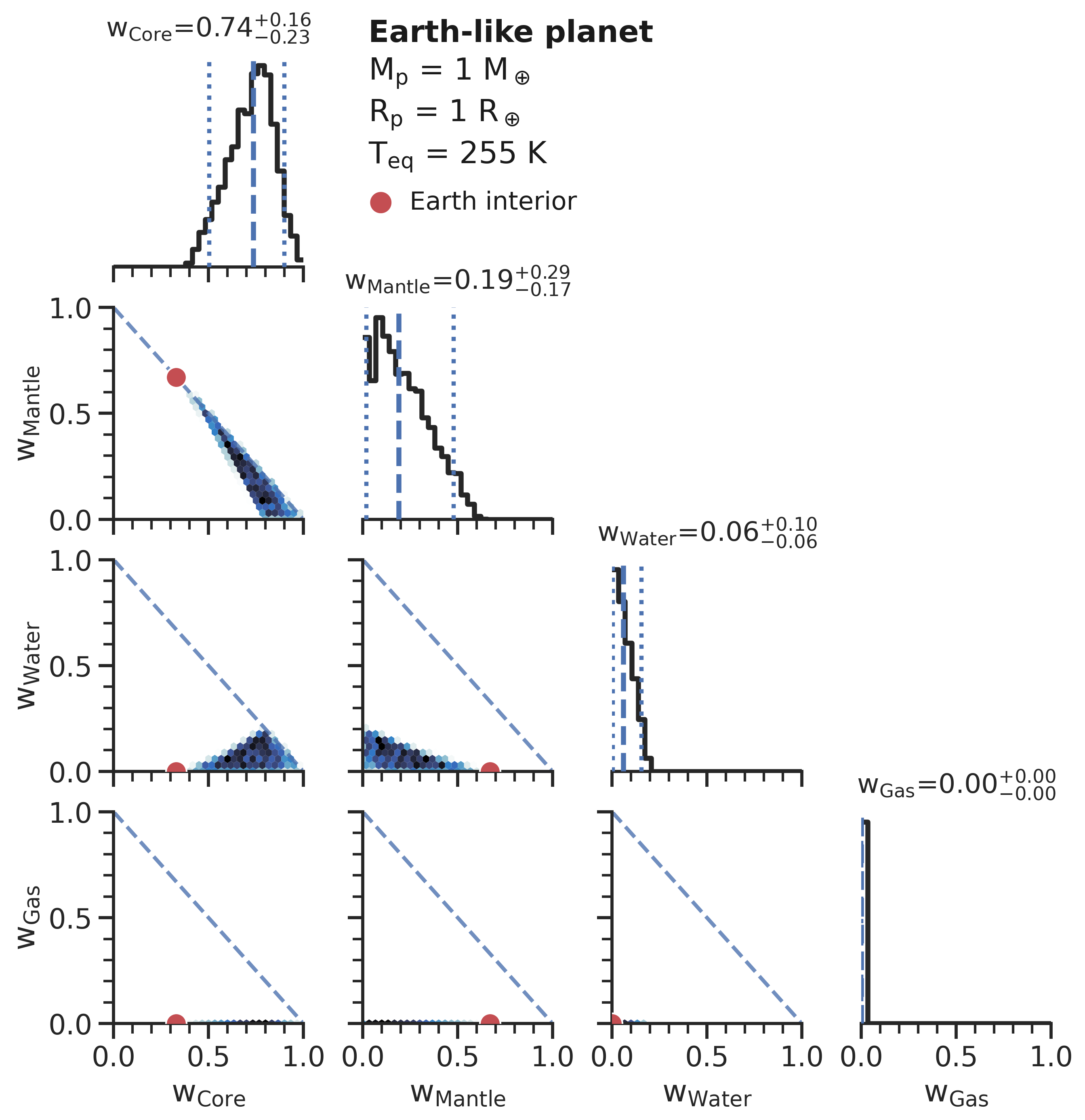}
    \caption{Predicted mass fraction of interior layers for an Earth-like planet with \SI{1}{\Me} and \SI{1}{\Re}. The red circle corresponds to Earth's true interior structure. Due to the presence of lighter elements in Earth's core, the actual core mass of Earth 33\% is slightly lower than what is predicted by ExoMDN for a solution with no water and atmosphere (39\%). The diagonal plots show the marginal distributions of each layer, with the blue dashed lines marking the median value and the dotted lines the 5th and 95th percentiles.}
    \label{fig:earth_mf}
\end{figure}

\begin{figure}[htb!]
    \centering
    \includegraphics[width=\linewidth]{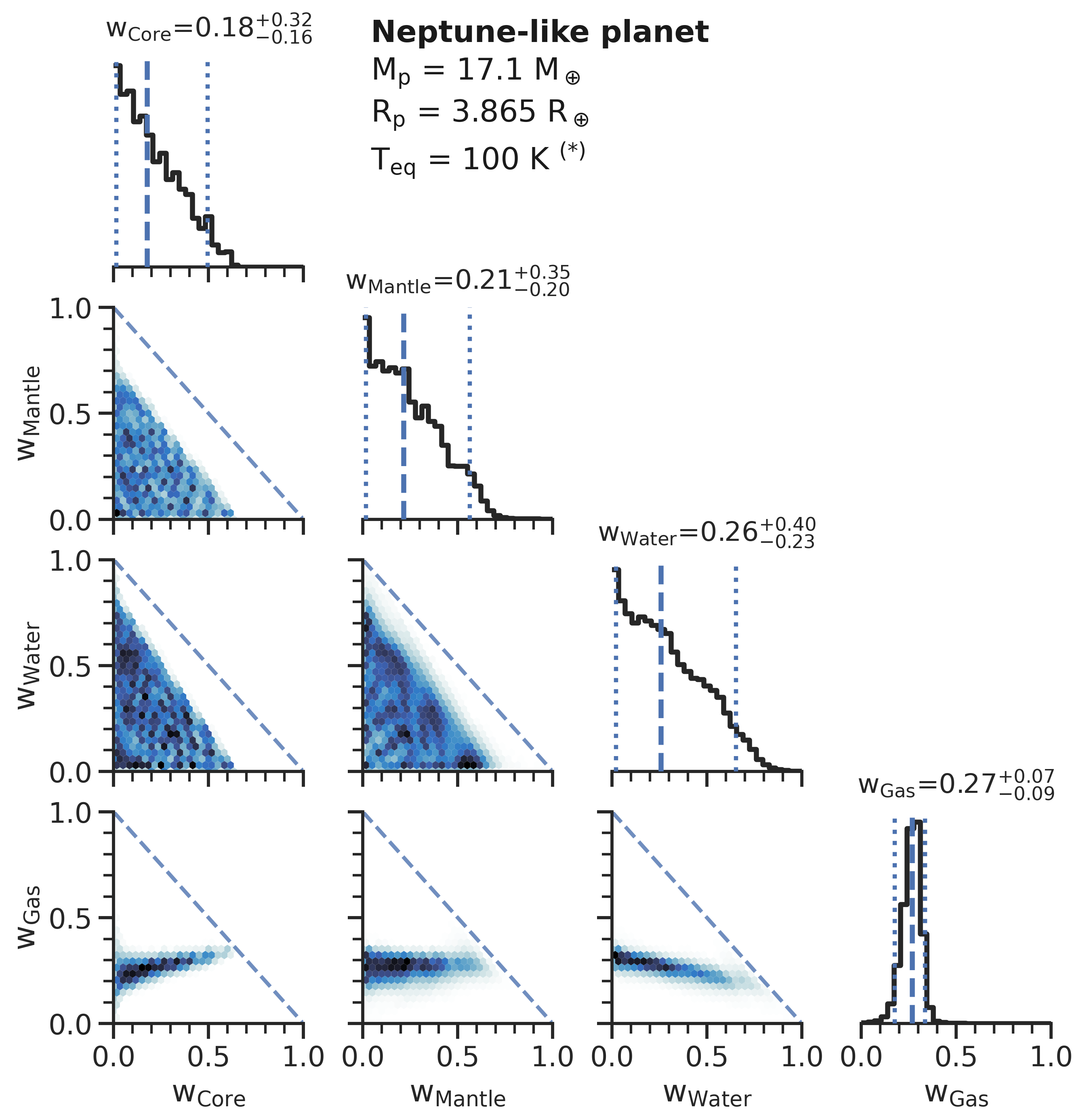}
    \caption{Predicted mass fraction of interior layers for a Neptune-like planet with \SI{17.1}{\Me} and \SI{3.865}{\Re}. The diagonal plots show the marginal distributions of each layer, with the blue dashed lines marking the median value and the dotted lines the 5th and 95th percentiles. $^{(*)}$Instead of Neptune's equilibrium temperature of \SI{51}{K}, a value of \SI{100}{K} was used to be in line with the parameter range of the training data.}
    \label{fig:neptune_mf}
\end{figure}

\begin{figure}[htb!]
    \centering
    \includegraphics[width=\linewidth]{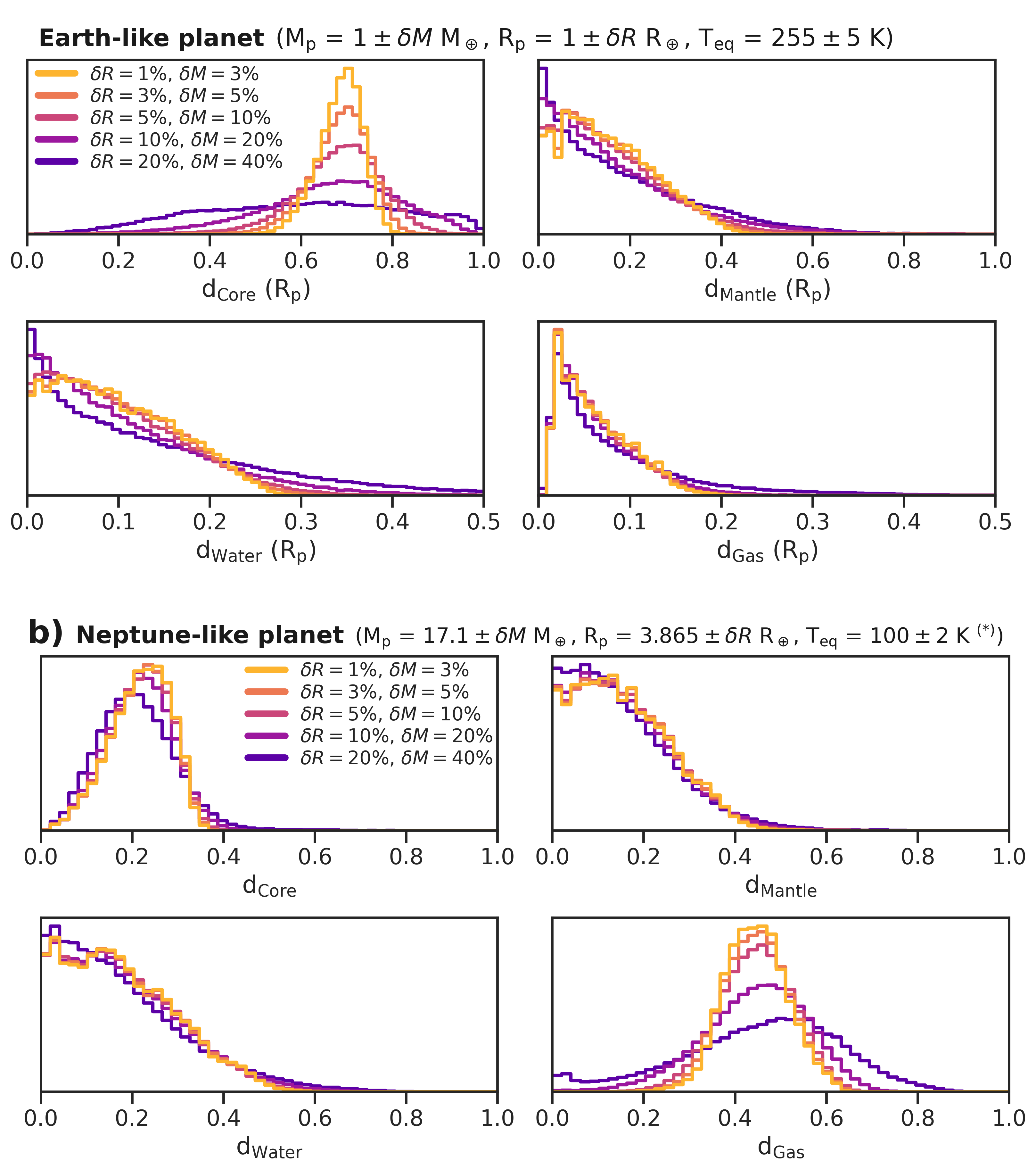}
    \caption{Effect of radius ($\delta R$) and mass ($\delta M$) uncertainty on the ability to constrain the interior for Earth (a) and Neptune (b) analogs. Each panel shows the marginal distributions for each interior layer with increasing amounts of radius and mass uncertainty. A $T\_{eq}$ uncertainty of 2\% has been assumed for both planets. $^{(*)}$Instead of Neptune's equilibrium temperature of \SI{51}{K}, a value of \SI{100}{K} was used to be in line with the parameter range of the training data.}
    \label{fig:mr_err}
\end{figure}

\begin{figure}[htb!]
    \centering
    \includegraphics[width=\linewidth]{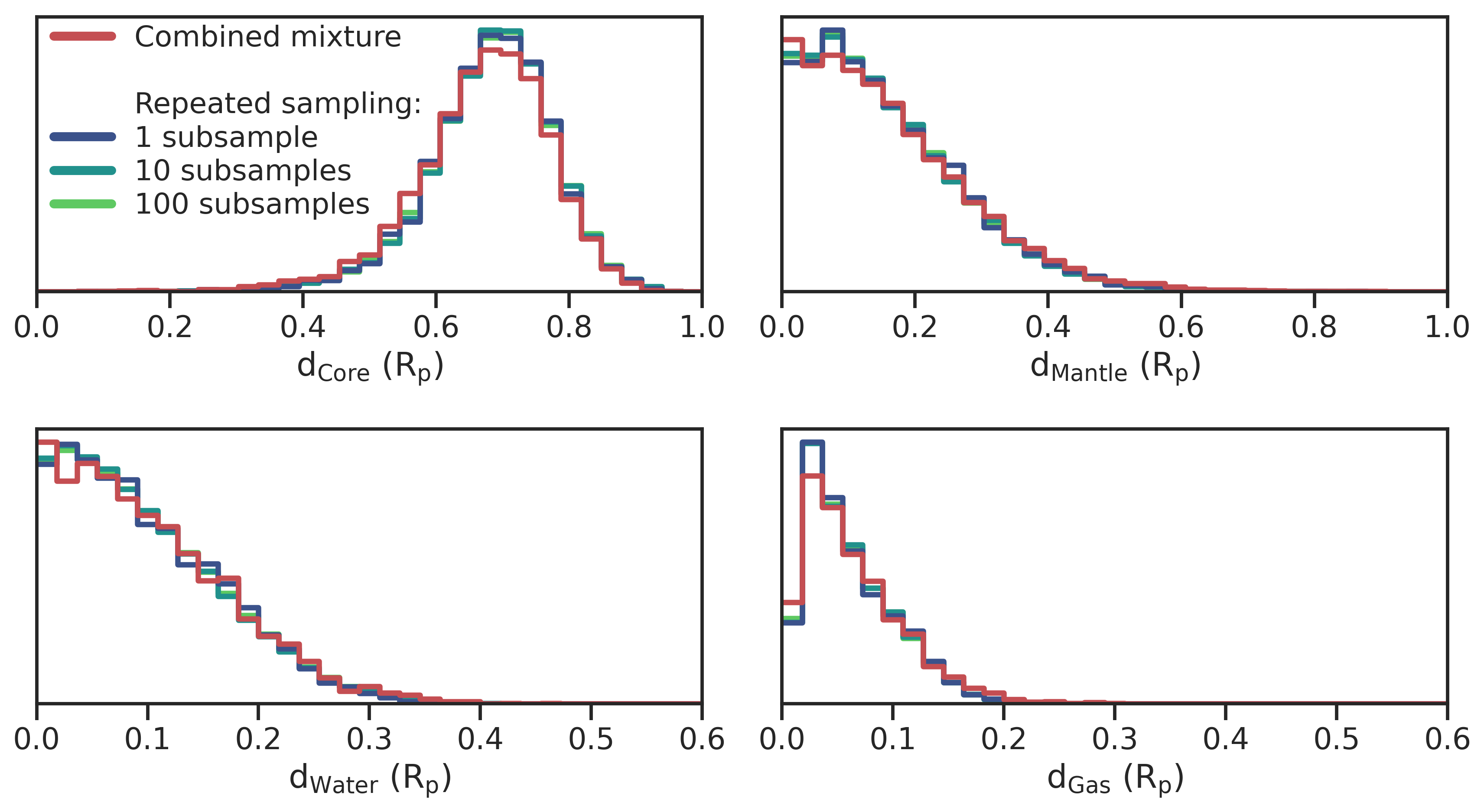}
    \caption{Comparison of the two possible approaches to incorporate measurement uncertainties (Sec. \ref{sec:uncertainty})  for an Earth-like planet with 5\% radius and 10\% mass uncertainty. 5000 posterior distributions were predicted from random mass, radius, and $T\_{eq}$ inputs within the uncertainties. The red line shows the predicted thickness of each interior layer obtained by first summing up all 5000 posterior distributions and then taking 5000 random samples from the combined mixture, while the dark blue, light blue, and green lines show those obtained by first taking 1, 10, and 100 samples, respectively, from each of the 5000 posteriors and then combining the samples.}
    \label{fig:errorsampling_comparison}
\end{figure}

\begin{figure}[htb!]
    \centering
    \includegraphics[width=\linewidth]{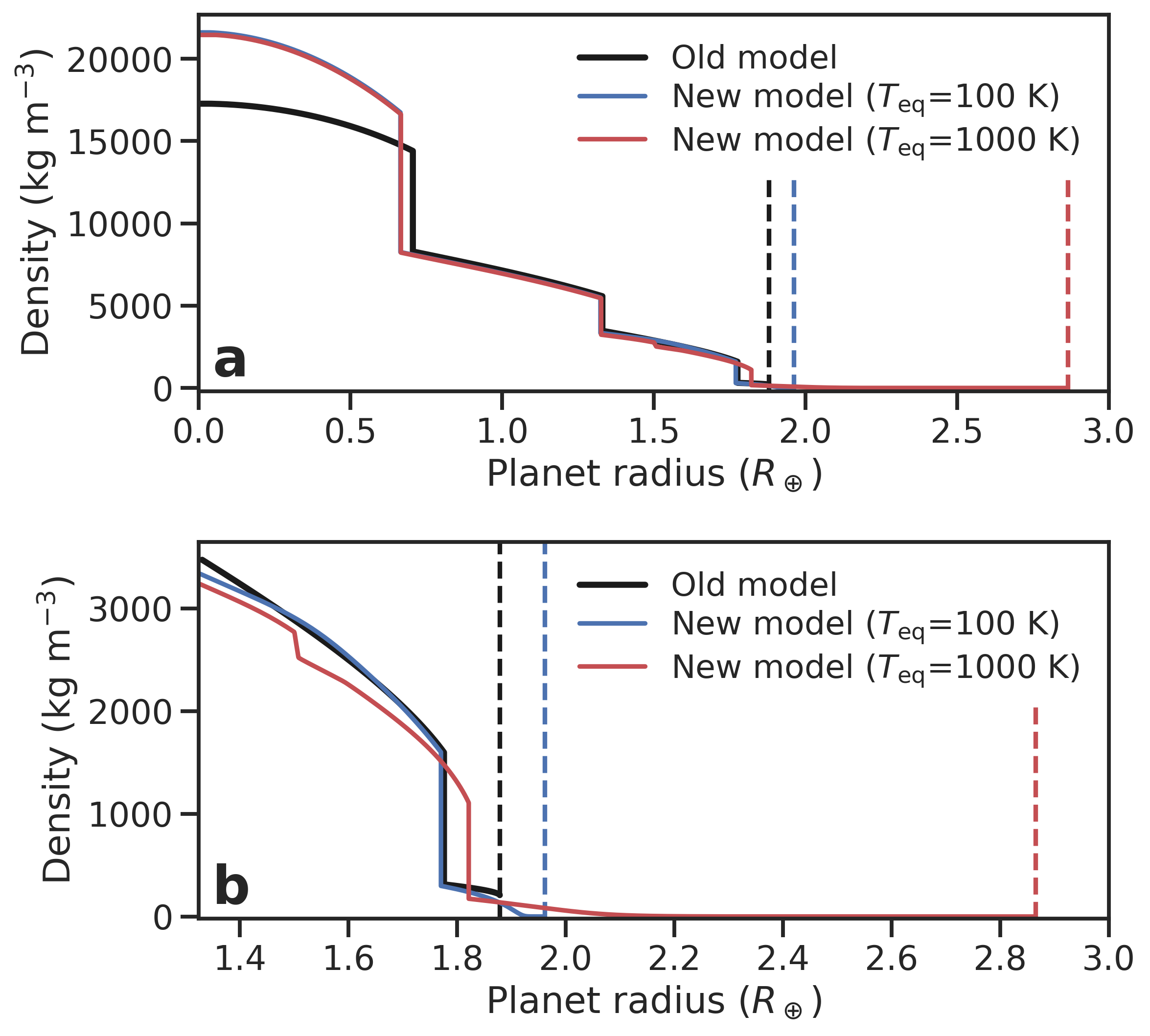}
    \caption{Illustration of the differences in interior models between the previous work \citep[][black line]{baumeister2020MachinelearningInference} and this work (blue and red lines, for two different equilibrium temperatures $T\_{eq}=100 K$ and $T\_{eq}=1000 K$, respectively). The figure shows density profiles of a representative \SI{5}{\Me} planet with $w\_{Core}=0.2$, $w\_{Mantle}=0.49$, $w\_{Water}=0.3$, and $w\_{Gas}=0.01$ (Panel a). Panel b shows a zoomed-in view of only the water and atmosphere layers. The dashed lines mark the respective planet radii.}
    \label{fig:modelcomp}
\end{figure}
\end{appendix}
\end{document}